\documentclass[fleqn,usenatbib]{mnras}
\usepackage{newtxtext,newtxmath}
\usepackage[T1]{fontenc}
\usepackage{threeparttable}
\DeclareRobustCommand{\VAN}[3]{#2}
\let\VANthebibliography\thebibliography
\def\thebibliography{\DeclareRobustCommand{\VAN}[3]{##3}\VANthebibliography}

\usepackage{graphicx}	
\usepackage{amsmath}	
\usepackage{amssymb}	

\usepackage{caption}

\title[Hydrodynamic Evolution of Sgr A East]{Hydrodynamic Evolution of Sgr A East: The Imprint of A Supernova Remnant in the Galactic Center}

\author[Mengfei Zhang, et al.]{
Mengfei Zhang,$^{1,2}$\thanks{E-mail: zmf@nju.edu.cn}
Zhiyuan Li,$^{1,2}$\thanks{E-mail: lizy@nju.edu.cn}
Ziqian Hua$^{1,2}$
Mark R. Morris$^{3}$
\\
$^{1}$School of Astronomy and Space Science, Nanjing University, Nanjing 210023, China\\
$^{2}$Key Laboratory of Modern Astronomy and Astrophysics (Nanjing University), Ministry of Education, Nanjing 210023, China\\
$^{3}$Department of Physics and Astronomy, University of California, Los Angeles, CA 90095, USA
}

\begin{document}

\label{firstpage}
\pagerange{\pageref{firstpage}--\pageref{lastpage}}
\maketitle






\begin{abstract}
We perform three-dimensional numerical simulations to study the hydrodynamic evolution of Sgr A East, the only known supernova remnant (SNR) in the center of our Galaxy, to infer its debated progenitor SN type and its potential impact on the Galactic center environment. Three sets of simulations are performed, each of which represents a represent a certain type of SN explosion (SN Iax, SN Ia or core-collapse SN) expanding against a nuclear outflow of hot gas driven by massive stars, whose thermodynamical properties have been well established by previous work and fixed in the simulations.
All three simulations can simultaneously roughly reproduce the extent of Sgr A East and the position and morphology of an arc-shaped thermal X-ray feature, known as the ``ridge''.  Confirming previous work, our simulations show that the ridge is the manifestation of a strong collision between the expanding SN ejecta and the nuclear outflow.
The simulation of the core-collapse SN, with an assumed explosion energy of $5\times10^{50}$ erg and an ejecta mass of 10 M$_\odot$, can well match the X-ray flux of the ridge,
whereas the simulations of the SN Iax and SN Ia explosions underpredict its X-ray emission, due to a smaller ejecta mass. All three simulations constrain the age of Sgr A East to be $\lesssim$1500 yr and predict that the ridge should fade out over the next few hundred years. We address the implications of these results for our understanding of the Galactic center environment.
\end{abstract}


\section{Introduction}\label{sec:intro}

Thanks to its proximity, the Galactic center offers a unique opportunity for disentangling the interplay between a supermassive black hole (SMBH) and its environment in unparalleled detail.
The Galactic center SMBH, commonly known as Sgr A*, is currently extremely underluminous for its well-determined gravitational mass of $4\times10^6\rm~M_\odot$ \citep{2019ApJ...882L..27D, 2020A&A...636L...5G}.
Surrounding and physically coupled to Sgr A* is the Nuclear Star Cluster (NSC), which mainly consists of $\sim10^7$ low-mass, old stars occupying the central parsecs \citep{2017MNRAS.466.4040F}.
At the center of the NSC, lying within $\sim$0.5 pc of Sgr A*, is  the young nuclear cluster (YNC) containing more than a hundred massive stars.
The YNC is the result of the latest episode ($\sim$ 4--6 Myr ago) of star formation around Sgr A* \citep{Genzel2010,2013ApJ...764..155L}.
Most prominent among the YNC are $\sim$30 Wolf-Rayet (WR) stars, each producing strong winds at a mass loss rate of $\sim10^{-5}\rm~M_\odot~yr^{-1}$ and a terminal velocity of $\sim1000\rm~km~s^{-1}$ \citep{Martins2007}.
The mutual collisions
of these stellar winds create strong shocks, leading to rapid thermalization of the wind kinetic energy (e.g., as observed in the case of IRS 13E; \citealp{2020ApJ...897..135Z,2020MNRAS.492.2481W}).
This ultimately results in a complex network of hot gas with temperatures of $\sim10^{7}$ K pervading the colliding wind zone, as demonstrated by dedicated hydrodynamic simulations \citep{2004ApJ...604..662R, Cuadra2006,Cuadra2008, 2017MNRAS.464.4958R,2018MNRAS.478.3544R}.
It is generally accepted that Sgr A* is currently mainly fed by the shocked winds, i.e., a {\it Bondi-like} accretion, at an extremely sub-Eddington yet significantly variable level \citep{1997ApJ...488L.149C,2005MNRAS.360L..55C,2020ApJ...888L...2C, 2020MNRAS.492.3272R, 2020ApJ...896L...6R}.

Outside the effective Bondi radius of $\lesssim0.1$ pc, the hot gas is not gravitationally bound, but ultimately forms an outflow \citep{2004ApJ...613..322Q, Cuadra2006}.
This outflow, hereafter dubbed the {\it nuclear outflow}, carries away the bulk mass, momentum and energy released by the massive stars, thereby providing an effective shielding for Sgr A* from being heavily fed by the circumnuclear medium. Therefore, a quasi-steady inflow-outflow system is established in the central parsecs around Sgr A* \citep{Shcherbakov2010}, which, in the absence of violent outbursts from Sgr A* itself, should be only mildly modulated by the orbital motion of the WR stars on a timescale of $\lesssim1000$ yr.

One expects that this quasi-steady state will be eventually broken when one or more of the WR stars evolves to a core-collapse supernova (CCSN), the kinetic energy and momentum of which can greatly reshape the circumnuclear gas.
Theoretical studies show that, depending on the strength of the SN and its distance from the SMBH, the SN shock wave may sweep over the central volume otherwise shielded by the stellar winds \citep{Yalinewich2017} and may pump gas into the close vicinity of the SMBH, potentially boosting its accretion level \citep{ 2020A&A...644A..72P,mnras3723}.

In fact, strong evidence exists for a SN shock currently encroaching upon the vicinity of Sgr A*.
High-resolution {\it Chandra} observations have revealed an arc-shaped feature of diffuse X-ray emission to the east (in the sense of Galactic coordinates) of Sgr A* \citep{2002ApJ...570..671M}.
This feature, highlighted in Figure~\ref{fig:obs} and hereafter referred to as the ``X-ray ridge'' \citep{Rockefeller2005}, spans a radial range of 0.36--0.6 pc from Sgr A* and an azimuth of $\sim$ \textit{$170^{\circ}$}.
\citet{Rockefeller2005} proposed that the X-ray ridge originates from an ongoing collision between the cumulative winds of massive stars around Sgr A* and the forward shock of Sgr A East, the prominent shell-like, non-thermal radio source that is widely believed to be a supernova remnant (SNR;  \citealp{1983A&A...122..143E, 2002ApJ...570..671M}).
\citet{Rockefeller2005} and \citet{2006ApJ...638..786F} verified this scenario using smoothed particle hydrodynamic (SPH) simulations, finding a reasonable agreement with {\it Chandra} observations on the morphology and X-ray luminosity of the ridge and inferring a dynamical age $\lesssim2000$ yr for Sgr A East.

Sgr A East certainly plays an active and unique role in transforming the immediate environment of Sgr A*.
In this work, we revisit the evolution of Sgr A East and its interaction with the nuclear outflow using hydrodynamic simulations.
We are motivated by several recent developments.
The first motivation is related to the type of SN explosion that created Sgr A East.
For decades since the identification of Sgr A East as a SNR, a core-collapse progenitor has been favored. This is chiefly owing to (i) the fact that numerous massive stars exist in the Galactic center,
(ii) a high metal abundance inferred from the X-ray spectrum of Sgr A East, presumably dominated by the reverse shock-heated SN ejecta  \citep{2002ApJ...570..671M,2004MNRAS.350..129S}, and (iii) a candidate neutron star, known as the ``Cannonball'' \citep{Park2005}, whose current position and measured proper motion suggest a physical association with Sgr A East \citep{2013ApJ...777..146Z}.
However, this consensus was recently challenged by \citet{2021ApJ...908...31Z}, who, based on an updated X-ray spectroscopic measurement of metal abundances and abundance ratios of Sgr A East and a detailed comparison with modern SN nucleosynthesis models, favored a Type Iax supernova (SN Iax).
Compared to CCSNe, a SN Iax is characterized by a relatively low explosion energy and a small ejecta mass, which could strongly affect the subsequent hydrodynamic evolution. Hence it is timely to investigate the hydrodynamic evolution of a SN Iax in the vicinity of Sgr A* and to contrast it with the case of a CCSN, as was assumed by \citet{Rockefeller2005}.

A second motivation arises from significant recent progress that has led to a good understanding of the thermodynamic properties of the nuclear outflow and the gravitational potential around Sgr A* (see details in the following sections), thanks to dedicated multi-wavelength observations and comprehensive numerical simulations conducted over the past two decades.
This facilitates the use of numerical simulations to follow the hydrodynamic evolution of different types of SNe in the vicinity of Sgr A*, which in turn allows us to place an independent constraint on the explosive progenitor by confronting the key observed properties of Sgr A East. Last but not least, a deliberated case study of the currently only known SNR in the vicinity of Sgr A* is highly complementary to the theoretical investigations of \citet{Yalinewich2017}, \citet{2020A&A...644A..72P} and \citet{mnras3723}, which survey the general behavior of circumnuclear SNRs.

The remainder of this paper is organized as follows.
We outline the key observational constraints in Section~\ref{sec:obs} and describe the setup of the hydrodynamic (HD) simulations in Section \ref{sec:sim}.
The simulation results are presented in Section \ref{sec:res}.
Implications of the results as well as caveats of our simulations are discussed in Section \ref{sec:dis}.
We assume a canonical distance of 8 kpc to Sgr A* ($1\arcsec$ corresponds to 0.039 pc) in this work.

\section{Observational Constraints}
\label{sec:obs}

\begin{figure*}
\includegraphics[width=\textwidth]{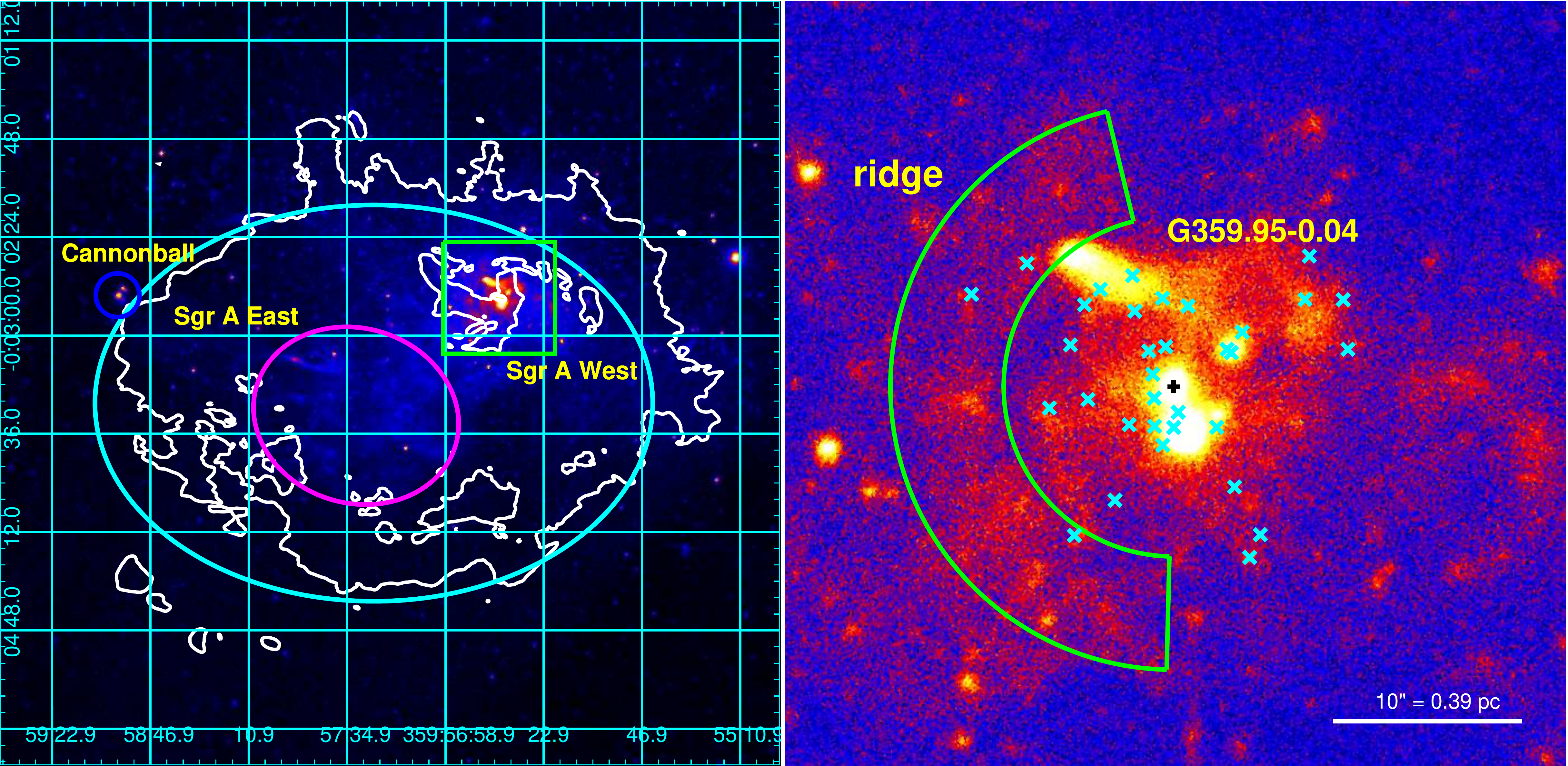}
\caption{{\it Left:} A {\it Chandra} X-ray image of the inner few parsec region of the Galactic center (from \citealp{2021ApJ...908...31Z}), overlaid by VLA 5.5 GHz continuum intensity contours (from \citealp{2013ApJ...777..146Z}).
The cyan ellipse delineates the radio shell of Sgr A East, while the magenta ellipse highlights the X-ray-bright interior. {\it Right:} A zoom-in view of the central 1.6 pc$\times$1.6 pc region outlined by the green rectangle in the left panel. The positions of known WR stars, which are responsible for the nuclear outflow, are marked by the `x' symbols, while Sgr A* is marked by the `+' symbol.
The X-ray ridge is delineated by the green wedge. Galactic coordinates are adopted in both panels.}
\label{fig:obs}
\end{figure*}

We consider three key observed properties of Sgr A East as robust constraints on the modelling of its evolution.
The first is the current physical size of Sgr A East, which is essentially defined by the extent of its radio shell.
In the left panel of Figure~\ref{fig:obs}, we plot radio continuum intensity contours of the Sgr A complex, obtained from the VLA 5.5 GHz map of \citet{2013ApJ...777..146Z}, against the X-ray emission from the same region, as seen by a deep {\it Chandra} image originally presented in \citet{2020ApJ...897..135Z} and \citet{2021ApJ...908...31Z}.
We approximate the radio shell with a cyan ellipse of a size of $\rm 8.2~pc \times5.8~pc$, which will be used to guide the SN forward shock in the HD simulations.
We note that the radio shell has a clear elliptical morphology on its eastern side, whereas a precise determination of its western extent is complicated by the presence of Sgr A West \citep{2005ApJ...620..287H} and the presumed interaction between the forward shock and the nuclear outflow.
Despite this uncertainty, the geometric center of the radio shell can be readily inferred, which is assumed to be the SN explosion center in this work, though \cite{Lau2015} questioned this assumption (see further discussion in Section \ref{subsec:ridge}).
However, this assumption is reinforced by the second observational aspect, i.e., the X-ray-bright core of Sgr A East, which is approximately delineated by a magenta ellipse in Figure~\ref{fig:obs}.
The X-ray spectra of this region reveal strong metal lines, especially from highly-ionized Fe, which are generally thought to be due to the SN ejecta \citep{2002ApJ...570..671M,2004MNRAS.350..129S,Park2005,2007PASJ...59S.237K,2019PASJ...71...52O,2021ApJ...908...31Z}.
The third and the most peculiar aspect is the presence of the X-ray ridge, as highlighted in the right panel of Figure~\ref{fig:obs}. 
Its current position, morphology and surface brightness should provide a strong constraint on the properties of the incoming SN shock, given the thermodynamics properties of the nuclear outflow (see Section~\ref{subsec:sw}).
A satisfactory HD simulation is expected to reproduce the above three observational aspects.

\section{Simulation}\label{sec:sim}

We use the publicly available, modular HD code \textit{PLUTO}\footnote{http://plutocode.ph.unito.it/} \citep{Mignone2007, Mignone2012} to simulate the early evolution of Sgr A East.
This grid-based HD code, with a second-order Runge–Kutta time integrator and a Harten-Lax-van Leer Riemann solver for middle contact discontinuities, is well suited for simulating the interaction between the SN shock and the nuclear outflow induced by the WR stars surrounding Sgr A*.

\subsection{Basic configuration}\label{subsec:config}

The HD simulations for the SNR evolution are performed on a three-dimensional (3D) Cartesian frame with an Adaptive Mesh Refinement (AMR) grid.
The AMR is desired due to the substantial dynamic range involved: Sgr A East currently has a physical size of 8.2 pc $\times$ 5.8 pc, the nuclear outflow is launched from a region spanning a few tenths of a parsec around Sgr A*, and the interaction between the SN shock and the nuclear outflow is resolved by {\it Chandra} on a scale of $\lesssim$0.04 pc.
Moreover, the interaction zone is highly dynamic, which precludes the use of a static mesh refinement.
In the AMR setup, we use a base grid (level 0) of $128^3$ and three refinement levels (levels 1, 2 and 3) having an equivalent grid of $256^3$, $512^3$ and $1024^3$,
respectively.
Grid refinement takes place following a substantial change in the total energy density.
The physical volume of the simulation box is set as $\rm 10^3~pc^3$ to cover the entire radio shell of Sgr A East.
Hence the linear resolution of levels 0, 1, 2 and 3 is $\sim$ 0.08, 0.04. 0.02 and 0.01 pc pixel$^{-1}$, respectively.
In addition, we perform seed simulations for the nuclear outflow (Section~\ref{subsec:sw}) on a 3D Cartesian frame with a fixed grid of $512^3$, which are then expanded to $1024^3$ by using linear interpolation.

We set the $z$-axis to be perpendicular to the Galactic disk (north as positive), the $y$-axis to run along decreasing Galactic longitude, and the $x$-axis to be parallel to the line-of-sight (the observer at the positive side).
The three-dimensional offset of the SN explosion center with respect to Sgr A* is not known {\it a priori} and cannot be fully determined by observations.
Nevertheless, guided by the roughly elliptical shape of the radio shell (Figure~\ref{fig:obs}) and the commonly accepted fact that the bulk of Sgr A East is located behind Sgr A* along the line-of-sight \citep[e.g.,][]{2005ApJ...620..287H}, we estimate the position of the explosion center and set it as the origin of the simulation box.
This places Sgr A* at $(x, y, z)_{\rm BH}$ = (1.0, 1.8, 1.5) pc.
The uncertainty of the explosion center could be a few tenths of a parsec in the $y-z$ plane and somewhat larger in the $x$-direction.
However, we note that the dynamical age of Sgr A East, which is primarily constrained by the projected size of the radio shell and the recent formation of the X-ray ridge (see Section~\ref{sec:res}), is insensitive to the line-of-sight position of the explosion center.
In Appendix~\ref{app:pos}, we present several test simulations that explore the effect of a different position of the explosion center.


The simulations neglect magnetic fields, which are prevalent in the Galactic center \citep{2020A&A...644A..71G} and can in principle affect the dynamics of SN-driven flows \citep{2021ApJ...913...68Z}. However, both the strength and orientation of the magnetic field in the innermost parsecs of the GC are currently not well constrained. Hence in the present study we assume that the dynamic evolution of Sgr A East is not significantly affected by the magnetic field.
In this case, the fluid evolution can be described by the ideal HD equations,
\begin{eqnarray}
\frac{\partial \rho}{\partial t} + \nabla \cdot (\rho \bf{v}) = \dot{\rho}_{\rm wind} + \dot{\rho}_{\rm SN}, \\
\frac{\partial (\rho\bf{v})}{\partial t}+\nabla \cdot\left[\rho\bf{vv}+\bf{1}p\right]^{T}=-\rho \nabla \Phi + \dot{\Pi}_{\rm wind} + \dot{\Pi}_{\rm SN},\\
\frac{\partial E_{t}}{\partial t}+\nabla \cdot\left[\left(E_t+p+\rho \Phi\right)\bf{v}\right] = -\frac{\partial\left( \rho \Phi\right)}{\partial t} + \dot{E}_{\rm wind} + \dot{E}_{\rm SN},
\end{eqnarray}
where $\rho$ is the mass density, $p$ the thermal pressure, $\bf{v}$ the velocity, $\bf{1}$ the dyadic tensor, $\Phi$ the gravitational potential, and $E_t$ the total energy density defined as
\begin{eqnarray}
    E_t = \rho \epsilon + \frac{\rho\bf{v}^2}{2},
\end{eqnarray}
where $\epsilon$ is the internal energy.
We adopt an ideal equation of state, i.e., $\rho \epsilon = p/ (\Gamma -1)$, in which the ratio of specific heats $\Gamma$ = 5/3.
The source terms, ($\dot{\rho}_{\rm wind}, \dot{\Pi}_{\rm wind}, \dot{E}_{\rm wind})$ and ($\dot{\rho}_{\rm SN}, \dot{\Pi}_{\rm SN}, \dot{E}_{\rm SN})$, consist of the continuous input from the WR star winds and the instantaneous input from the SN explosion (see Section~\ref{subsec:sw} for details of implementation).

The gravitational potential in the simulation volume is assumed to be static and completely determined by the SMBH and the NSC.
A point mass of $4\times10^6\rm~M_{\odot}$ is taken to represent the SMBH.
For the NSC, we adopt a spherically symmetric distribution following
\citet[Equation 5 therein]{2015MNRAS.447..948C}, which has an enclosed mass of $\rm 6.5\times10^6~M_{\odot}$ within a radius of 5 pc.
Although a spheroidal model of the NSC can better explain the velocity dispersion outside the central 5 pc  \citep{2015MNRAS.447..948C}, a spherical model is sufficiently accurate for the stellar mass distribution within our simulation box of 10$^3$ pc$^3$.

The simulations  neglect viscosity but include thermal conduction and radiative cooling. For the latter, we adopt the TABULATED cooling function implemented in \textit{PLUTO}, which is generated by \textit{Cloudy} \citep{2017RMxAA..53..385F} for an optically thin plasma of a given elemental abundance.
We impose a minimum cooling temperature of 10$^4$ K, the exact choice of which does not significantly affect the HD evolution and the resultant X-ray emissivity.
We also adopt a floor for the gas number density ($n_{\rm floor} = 10^{-6}\rm~cm^{-3}$) to prevent numerical artifacts that may lead to negative values in this quantity.

\subsection{Nuclear outflow and supernova explosion}\label{subsec:sw}

The mutual interaction of the strong winds from $\sim$30 WR stars in the central half-parsec has been extensively studied by analytic models and numerical simulations \citep{2004ApJ...613..322Q,2004ApJ...604..662R,Cuadra2006,Cuadra2008,2018MNRAS.478.3544R}.
These studies demonstrate that wind-wind collisions lead to rapid thermalization of the wind kinetic energy, creating a complex network of hot gas that ultimately escapes in the form of a bulk outflow.
Since we are mainly interested in the interaction of this outflow with Sgr A East, our simulations neglect the details of wind-wind collision and the chaotic gas motions within the wind colliding zone. Instead, we implement a smooth and isotropic radial outflow by employing an internal boundary in {\it PLUTO}, which approximates the gas density, temperature and velocity profiles in a spherical shell between $r = 0.08 - 0.2$ pc following the results of \citet{2004ApJ...613..322Q} and \citet{2018MNRAS.478.3544R}:
$n \propto  r^{-1}\rm~cm^{-3}$, $T=1.5 r^{-0.56}\rm~keV$, $ v/v_{\rm out}=0.64{\rm log}(r)+0.62$, where $r$ is the radial coordinate from Sgr A* in units of pc, and the terminal velocity $v_{\rm out}=1000\rm~km~s^{-1}$.
The normalization of the density is such that the asymptotic mass loss rate  $\dot{M}_{\rm out} =  1\times10^{-3}\rm~M_{\odot}~yr^{-1}$, a value adopted by \citet{2018MNRAS.478.3544R} and compatible with the cumulative mass loss from the $\sim30$ WR stars.
In this shell, the resolution is 0.01 pc pixel$^{-1}$, or 12 grid points within the internal boundary.
The velocity setting ensures that the gas can flow both inward and outward.
Gas within $r \lesssim 0.04$ pc is bound by the gravity of Sgr A*.
To mimic the SMBH accretion and avoid unphysical gas pileup, we reset the gas density of the innermost pixels to a constant value ($10^{-3}\rm~cm^{-3}$) and the gas temperature to $10^4$ K after every time step.
Such a treatment does not affect the outward propagation of the nuclear outflow and its interaction with the SN shock.
To convert the gas number density into a mass density, we have assumed an element composition of ($X_{\odot}$=0, $Y_{\odot}$=0.96, $Z_{\odot}$=0.04), i.e., hydrogen-free and a three-times solar abundance for elements heavier than helium. This is consistent with the simulations of \citet{2018MNRAS.478.3544R} and motivated by the expectation that the WR star winds are hydrogen-depleted but carbon- and nitrogen-enriched.


The simulations begin with an initial uniform ambient density of $\rm0.1~cm^{-3}$ and a uniform temperature of 10$^6$ K.
After $\sim$ 1 Myr, the outflow propagates throughout the simulation box, resulting in distributions of density, temperature and radial velocity consistent with the overall profiles of \citet{2004ApJ...613..322Q} and
\citet{2018MNRAS.478.3544R}.
This quasi-steady environment for the subsequent explosion and evolution of the SN is completely determined by the assumed properties of the nuclear outflow.
 We note that this effectively neglects cold gas components, in particular, the circumnuclear disk (CND) of dense molecular gas, a ring-like structure roughly aligned with the Galactic plane and encompassing Sgr A West \citep{2012A&A...540A..50F, 2017ApJ...847....3H}. The potential effect of the CND is discussed in Section~\ref{subsec:disevo}.

Once the quasi-steady pre-SN conditions are established, the SN is implemented at the aforementioned explosion center in the form of injected energy ($E_{\rm SN}$) and ejecta mass ($M_{\rm SN}$).
Motion of the progenitor star relative to Sgr A* is neglected, which should be small (order 100~km~s$^{-1}$) compared to the SN shock velocity.
We consider three combinations of $E_{\rm SN}$ and $M_{\rm SN}$, which represent three possible types of supernova explosion.
The first type is SN Iax, as suggested by \citet{2021ApJ...908...31Z} for Sgr A East, which has
$E_{\rm SN}= 5\times10^{50}\rm~erg$ and $M_{\rm SN}=1.3\rm~M_{\odot}$.
The second type is SN Ia, for which we adopt $E_{\rm SN}= 1\times10^{51}\rm~erg$ and $M_{\rm SN}=1.3\rm~M_{\odot}$.
The last type is CCSN, for which we adopt $E_{\rm SN}= 5\times10^{50}\rm~erg$ and $M_{\rm SN}=10\rm~M_{\odot}$.
We note that SNe Iax can have a $E_{\rm SN}$ as low as $10^{50}\rm~erg$, while CCSNe can have $E_{\rm SN}$ up to $10^{51}\rm~erg$ and a wide range of $M_{\rm SN}$. The values chosen here are such that the SN Iax and CCSN simulations differ from the SN Ia simulation by just one of the two key parameters, which helps to discern the main effect of either explosion energy or ejecta mass.
In Appendix \ref{app:std}, we discuss a test simulation with $E_{\rm SN}= 5\times10^{51}\rm~erg$ and $M_{\rm SN}=10\rm~M_{\odot}$.
A more thorough exploration of the parameter space in $E_{\rm SN}$ and $M_{\rm SN}$ is deferred to a future work.

$E_{\rm SN}$ and $M_{\rm SN}$ are injected at time $t = 0$
within an effective radius of 0.2 pc (20 grid points) around the explosion center, following the analytic density, velocity and temperature profiles of \citet{2001ApJ...560..244B}.
The corresponding age of the SN at this radius is $\sim$ 10 yr \citep{Leahy2017a}, which is small compared to the evolution time up to the formation of the X-ray ridge (see Section \ref{sec:res}).
Specifically, for the SN Ia and Iax, we adopt a uniform core, a power-law ejecta density profile with a power-law index $n = 7$ and a constant ambient density profile with a power-law index $s = 0$, while for the CCSN, we adopt $n = 7$ and $s = 2$.
Our test simulations indicate that the results are insensitive to the exact values of $n$ and $s$ within reasonable ranges.
For simplicity, we assume a chemical composition for the SN ejecta same as for the nuclear outflow. Although in reality the ejecta could be significantly more metal-enriched, this choice has little effect on the HD evolution, nor does it significantly affect the synthetic X-ray map,  since the simulations show that the ejecta contribute little to the emission from the X-ray ridge (see Section~\ref{sec:res}).
This is further confirmed by our test simulation with a strongly metal-enriched ejecta ($Z_{\odot}$=0.08), which is detailed in Appendix~\ref{app:metal}.

\begin{figure}
\includegraphics[width=0.5\textwidth]{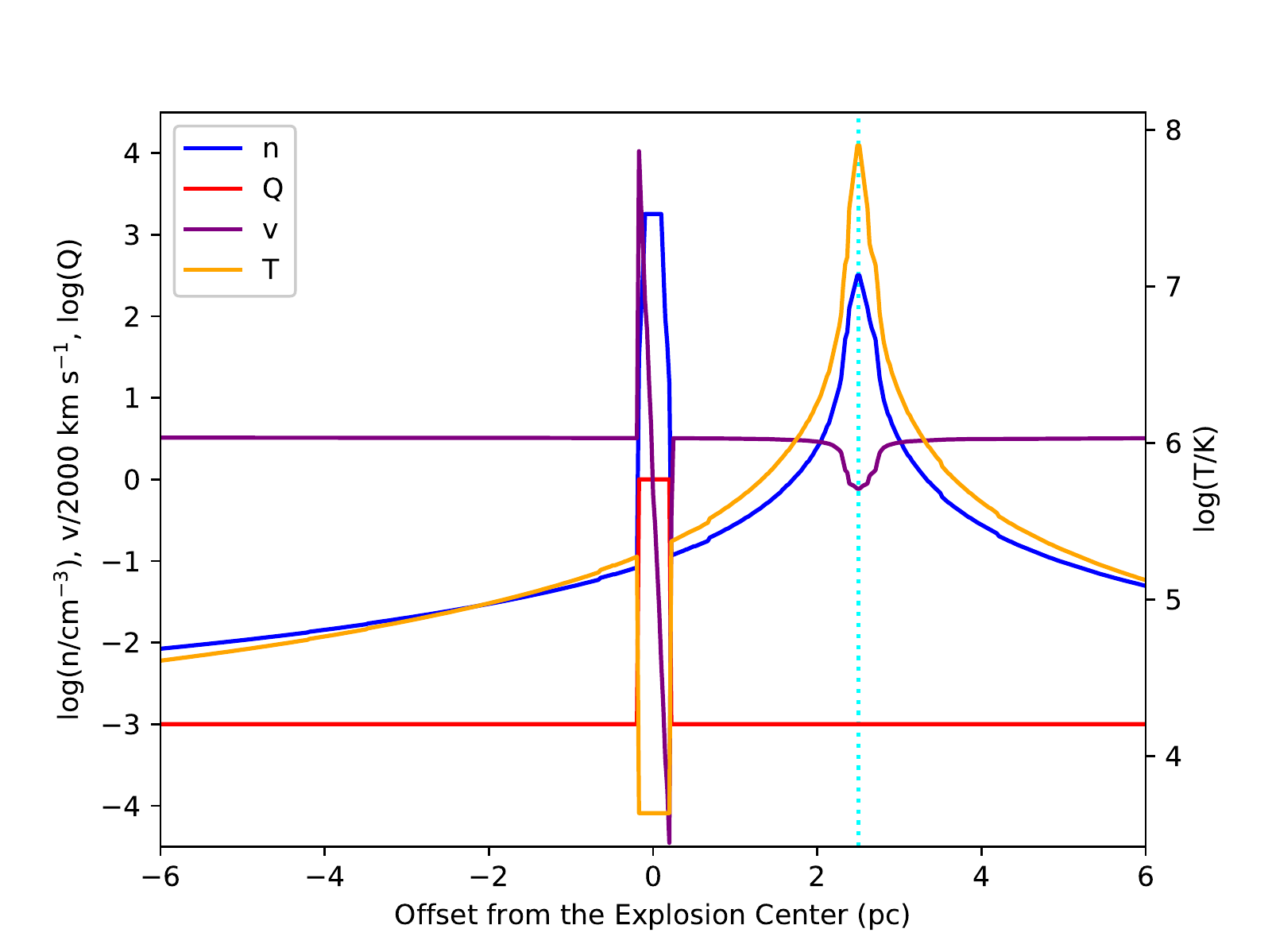}
\caption{One-dimensional profiles of density (blue), temperature (orange), velocity with respect to Sgr A* (purple; positive as away from Sgr A*) and ejecta tracer (red) along the line connecting the explosion center (zero-point) and Sgr A* (marked by a vertical dotted line), immediately after the SN explosion.
Here the tracer parameter assumes a lower bound of 0.001 for ease of visualization.
\label{fig:init}}
\end{figure}

Once injected, the ejecta is in contact with the nuclear outflow and the two fluids eventually partially mix with each other.
We introduce a tracer parameter, $Q$, which is evaluated at each pixel in the simulation and obeys a simple conservation law:
\begin{eqnarray}
      \frac{\partial (\rho Q)}{\partial t} + \nabla \cdot (\rho Q \bf{v}) = 0.
\label{eqn:tracer}
\end{eqnarray}
$Q$ has a value of 1 for pure SN ejecta and 0 for the unpolluted nuclear outflow, and values in between to indicate a mixed gas.
The physical conditions at $t = 0$ are illustrated in Figure \ref{fig:init}, which plots one-dimensional profiles of the gas density, temperature, velocity and ejecta tracer parameter along the line passing through the SN explosion center and Sgr A*.

\subsection{Simulation runs and synthetic X-ray emission}\label{subsec:runs}

With the above settings, we run a series of simulations.
Our fiducial simulation is denoted by run \textit{E5M1.3}, which represents an SN Iax explosion, as defined in Section~\ref{subsec:sw}, and serves as an illustration of the basic behavior of the SNR evolution (Section~\ref{subsec:fidu}).
The two other runs, \textit{E10M1.3} and \textit{E5M10}, represent the case of SN Ia and CCSN, respectively, and differ from the fiducial run in either the explosion energy or the ejecta mass.
The main parameters of the different runs are summarized in Table~\ref{table:para2}.
For all three runs, the total elapsed time since the injection of the supernova is set to be 2 kyr, which is roughly the time when the simulation box is almost filled by the ejecta in the fiducial run.
The time step is adaptive and ranges between $0.1-10$ yr.


\begin{table}
\center
\caption{Main parameters of the simulation runs}
\label{table:para2}
\begin{threeparttable}
\begin{tabular}{ccccc} 
\hline
Run & $E_{\rm SN}$ & $M_{\rm SN}$ & $t$ & $S_{2-8}$ \\
\hline
(1) & (2) & (3) & (4) & (5) \\
\hline
\textit{E5M1.3}   & 5  & 1.3  & 700 & 2.8 \\
\textit{E10M1.3}  & 10 & 1.3  & 500 & 4.9 \\
\textit{E5M10}    & 5  & 10   & 1400 & 6.0 \\
\hline
\end{tabular}
\begin{tablenotes}
\footnotesize
\item (1) Simulation run. (2) Explosion energy, in units of $10^{50}$ erg. (3) Ejecta mass, in units of $\rm M_\odot$. (4) The epoch (in units of yr) when the morphology of the synthetic X-ray ridge best matches the observed one (Section~\ref{sec:res}). (5) The predicted 2--8 keV surface brightness of the ridge at this epoch, in units of $10^{-15}\rm~erg~s^{-1}~cm^{-2}~arcsec^{-2}$.
\end{tablenotes}
\end{threeparttable}
\end{table}

To facilitate comparison with the observations, we generate synthetic X-ray intensity maps for selected snapshots of each simulation run.
The X-ray emissivity of an optically-thin thermal plasma in collisional ionization equilibrium (CIE; \citealp{2001ApJ...556L..91S}) is extracted from \textit{ATOMDB}\footnote{http://www.atomdb.org}, version 3.0.9, for which we adopt the elemental composition specified in Section~\ref{subsec:sw}.
The X-ray intensity maps are derived by integrating along the $x$-axis.
In calculating the synthetic X-ray emission, the electron temperature is assumed to be equal to the ion temperature.
This is justified by the estimated equipartition time of $\sim$100 yr between electrons and ions, due to Coulomb interaction \citep{1994ApJ...437..770M}, for density (10 cm$^{-3}$), ion temperature (10$^8$ K) and electron temperature (10$^6$ K) conditions characteristic of the X-ray ridge, which is comparable to the dynamical timescale.
On the other hand, the timescale for CIE is $\sim300(\rm cm^{-3}/n)$ yr for He and longer for heavier elements \citep{2010ApJ...718..583S}. Hence the assumption of CIE is not strictly valid, but this should have little effect on the resultant X-ray flux \citep{2009A&A...508..751S, 2019MNRAS.482.1602Z}.
An {\it a posteriori} comparison also finds a reasonable agreement between the emissivity-weighted ion temperature predicted by our simulations and the mean electron temperature inferred from the observed X-ray spectrum of the ridge \citep{2023arXiv230400920H}.

We note, in reality, the evolution of Sgr A East and its interaction with the nuclear outflow are determined by many factors, including the SN explosion energy, ejecta mass, the site and environment of the explosion, the nuclear outflow properties, etc.
With a primary focus on the formation of the X-ray ridge,
the exploration of the parameter space in this work is necessarily limited.
In particular, conditions of the medium at and near the SN explosion, such as the presence of dense cold cloudlets, may significantly affect the subsequent evolution.
However, the exact properties of such cloudlets are rather difficult, if at all possible, to accurately constrain. Thus, we do not consider them in this work (see further discussions in Section~\ref{subsec:disevo}), but note that
\citet{Rockefeller2005} and  \citet{2022A&A...668A.124E} have performed simulations to consider some specific conditions of the surrounding medium, which are complementary to our study here.
A more thorough exploration of the parameter space would be an interesting future work.

\section{Results}\label{sec:res}

In this section, we present the simulation results.
We first describe in detail the evolution of the SNR and the formation of the X-ray ridge in the fiducial run (Section~\ref{subsec:fidu}).
We then examine these processes in the two additional runs, one with a higher explosion energy (Section~\ref{subsec:he}) and the other with a higher ejecta mass (Section~\ref{subsec:cc}),  to illustrate how the change in one of two key parameters would affect the evolution of the SNR and the formation of the X-ray ridge.

\subsection{The run in the fiducial set}\label{subsec:fidu}
\begin{figure*}
\includegraphics[width=\textwidth]{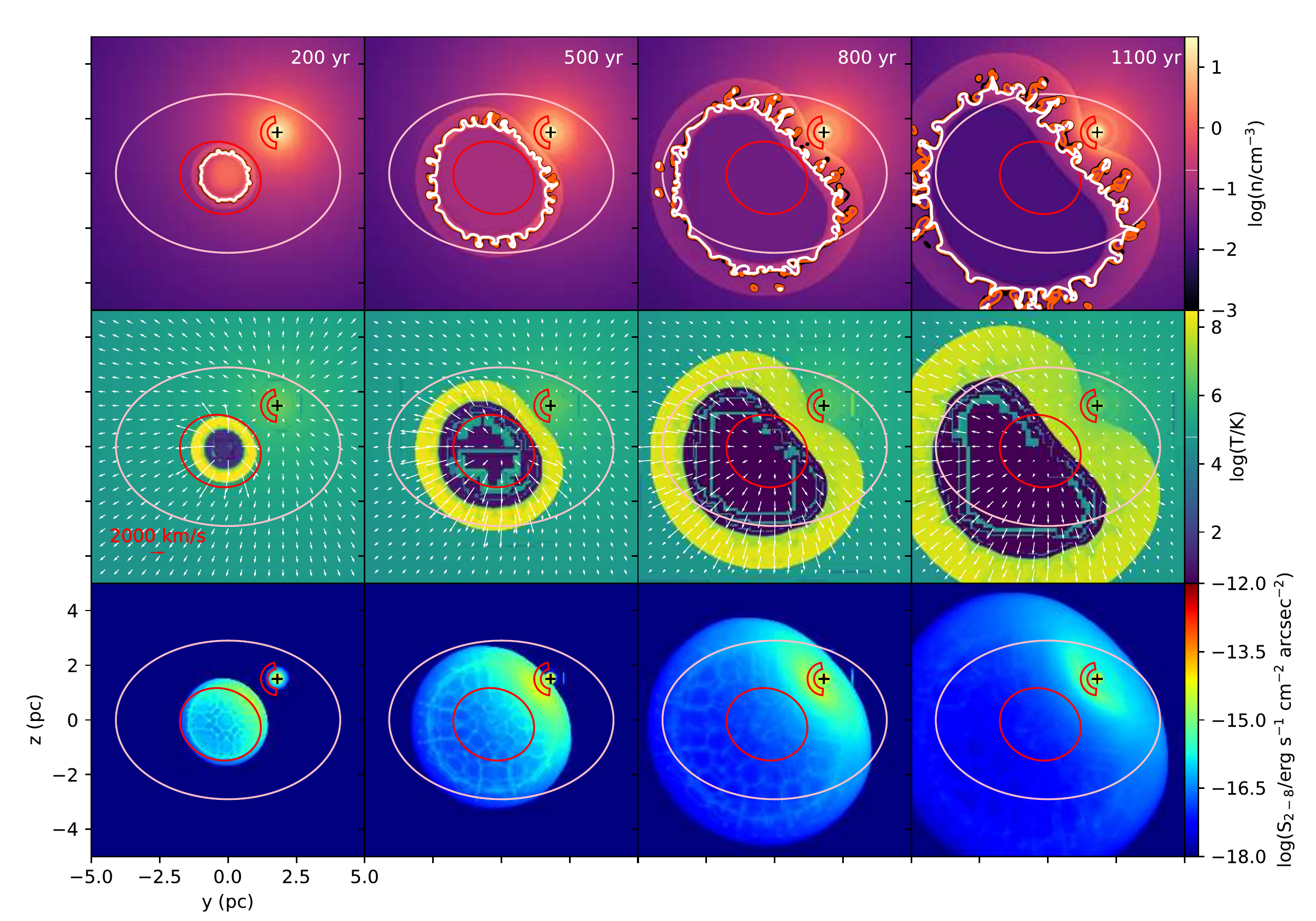}
\caption{Simulation run \textit{E5M1.3}. The {\it top} and {\it middle} panels show snapshots of density and temperature distributions in the $x=1$ plane at 200, 500, 800 and 1100 yr (from left to right) after the SN explosion at the origin,
while the {\it bottom} panels show the corresponding 2--8 keV X-ray surface brightness distribution.
Sgr A* is located at $(x, y, z) =(1.0, 1.8, 1.5)$ pc and marked by a black `+'.
The pink ellipse, red ellipse and red wedge outline the observed position of the radio shell,  X-ray-bright interior and X-ray ridge, respectively, as in Figure~\ref{fig:obs}.
The extent of the SN ejecta is outlined by the white, red and black contours in the top panels, which correspond to an ejecta tracer parameter of 0.9, 0.5 and 0.1, respectively. The white arrows in the middle panels indicate the projected velocity vector.
Some peculiar filaments seen in the central region of the temperature maps are artifacts due to the adaptive grid refinement.
\label{fig:5}}.
\end{figure*}

\begin{figure*}

\includegraphics[width=\textwidth]{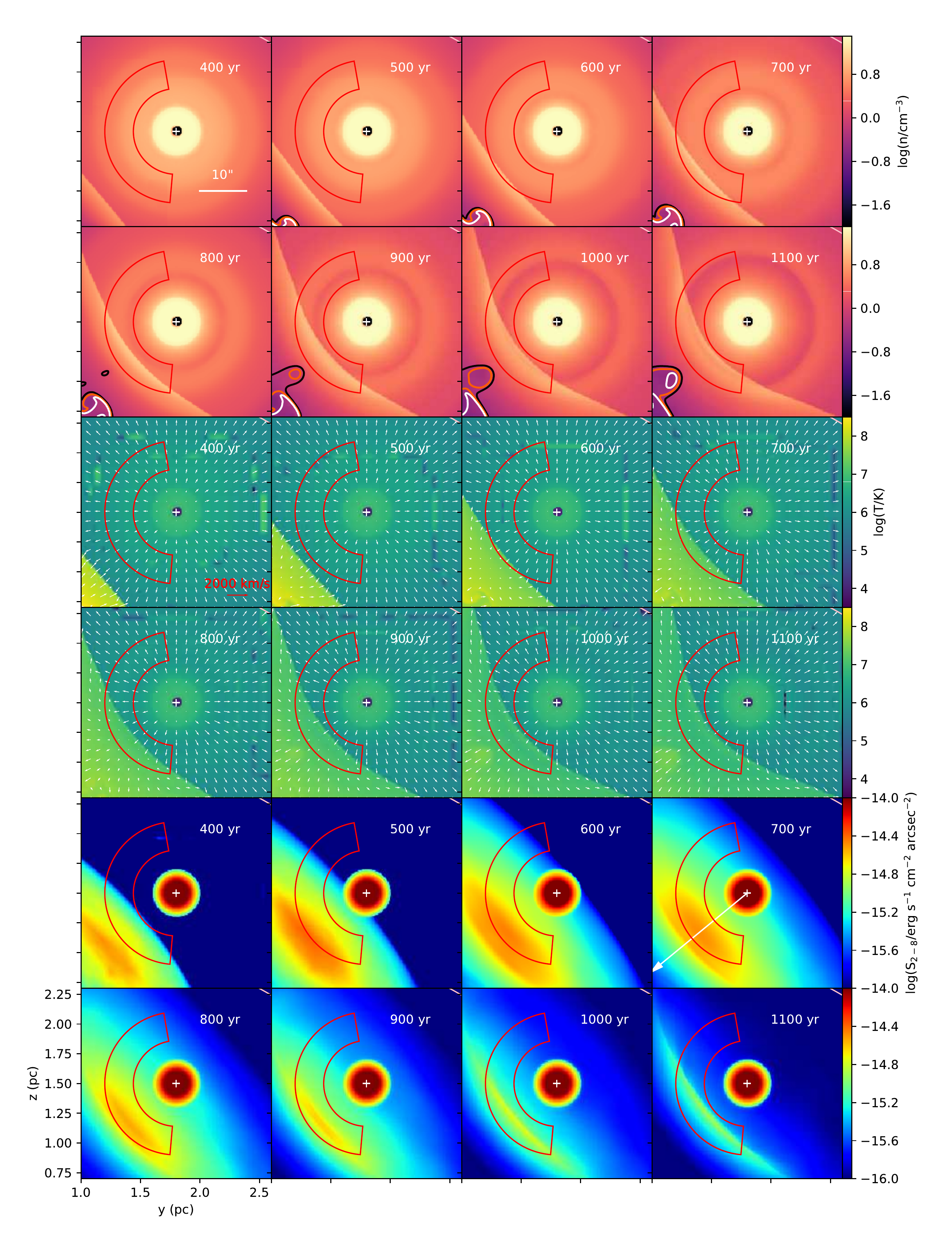}
\vskip-0.5cm
\caption{Similar to Figure~\ref{fig:5}, but zooming into a 1.6 pc $\times$ 1.6 pc region centered on Sgr A* and for snapshots from $t=$ 400 yr to 1100 yr in steps of 100 yr, in the simulation run \textit{E5M1.3}. The central dip is due to an artificial sink of gas to mimic accretion onto the SMBH. The white plus and red arc indicate Sgr A* and the X-ray ridge, respectively.
The white arrow points in the direction of the SN explosion center.
Color scales in the X-ray surface brightness maps (the bottom two rows) are adjusted to suppress the central outflow zone at radii of 0.2 pc from Sgr A* for an optimal illustration of the ridge.
\label{fig:He_tot_long}}.
\end{figure*}

The evolution of Sgr A East and its subsequent interaction with the nuclear outflow in run \textit{E5M1.3} are illustrated by the sequential snapshots presented in Figure \ref{fig:5} and \ref{fig:He_tot_long}.

In the top and middel panels of Figure \ref{fig:5}, we show the density and temperature distributions in the $x = 1$ pc plane (i.e., the plane across Sgr A*) at $t =$ 200, 500, 800 and 1100 yr after the SN explosion.
The artifacts shown in the panels can be attributed to the AMR, which arise primarily at the site of level transition.
The expanding ejecta drive a strong forward shock at an initial velocity of  $\rm \sim 10^4~km~s^{-1}$, compressing and heating up the ambient gas to temperatures $\gtrsim10^8$ K. In the meantime, a reverse shock develops back into the ejecta.
This results in the familiar morphology seen at $t = 200$ yr, where a nearly spherical shell of shock-heated hot gas encompasses the cold (due to adiabatic expansion), unshocked ejecta (delineated by the colored contours of the trace parameter in the density maps).
As it approaches Sgr A*, the forward shock is decelerated by the ram pressure of the nuclear outflow.
This results in a significant deviation of the shell from a round shape on the side facing the nuclear outflow by the time of $t= 500$ yr.
In fact, the ram pressure of the nuclear outflow is so strong that the forward shock can never penetrate into the region of $r \lesssim 0.3$ pc.
This is evident in the snapshot of $t= 800$ yr, where the forward shock bends around the central outflow zone.
In contrast, the forward shock has expanded almost freely on the opposite (i.e., southeast) side, reaching a radius comparable to the radio extent of Sgr A East (indicated by the pink ellipse) at $t= 800$ yr.
At later times, the forward shock continues to bend on the northwest side and progressively engulfs the central outflow zone, while on the southeast side it has moved out of the simulation box by the time of $t = 1100$ yr.
The bottom panels of Figure~\ref{fig:5} show the synthetic 2--8 keV X-ray surface brightness ($S_{2-8}$) distribution in the different epochs.
Generally, the synthetic X-ray morphology traces the expanding shell. High surface brightness ($S_{2-8} \gtrsim 10^{-15}\rm~erg~s^{-1}~cm^{-2}~arcsec^{-2}$) is found in the central outflow zone and the surrounding regions where the outflow interacts with the forward shock.
A low surface brightness cavity is coincident with the unshocked (hence cold) ejecta.
Notably, this is in sharp contrast with the observed X-ray-bright interior of Sgr A East (delineated by the red ellipse).
This discrepancy will be further addressed in Section~\ref{subsec:disevo}.

Figure \ref{fig:He_tot_long} provides a closeup view of the interaction between the forward shock and the nuclear outflow, in which the upper (middle) two rows depict the density (temperature) distribution in the central 1.6 pc $\times$ 1.6 pc region around Sgr A*, for eight consecutive snapshots from $t = 400$ yr to 1100 yr.
At $t = 400$ yr, the forward shock is approaching from the southeast, followed by the SN ejecta (indicated by the tracer contours in the density maps), but its interaction with the nuclear outflow is still weak at this time.
At $t = 500$ yr, the shock front comes closer to Sgr A*, and clearly its shape becomes flattened due to the ram pressure of the nuclear outflow.
The intrusion of the ejecta lags the forward shock.
Notably, the ejecta tracer contours exhibit finger-shaped patterns, which are understood as due to the Rayleigh-Taylor instability.
At $t = 600$ yr, the shock front reaches a distance of $\sim$0.5 pc from Sgr A* and starts to bend around an apex where the ram pressure of the nuclear outflow approximately equals that of the shock.
This effectively builds up a dynamic interaction zone  between the nuclear outflow and the SN ejecta, which is sandwiched by two standing shocks.

\begin{figure*}
\includegraphics[width=\textwidth]{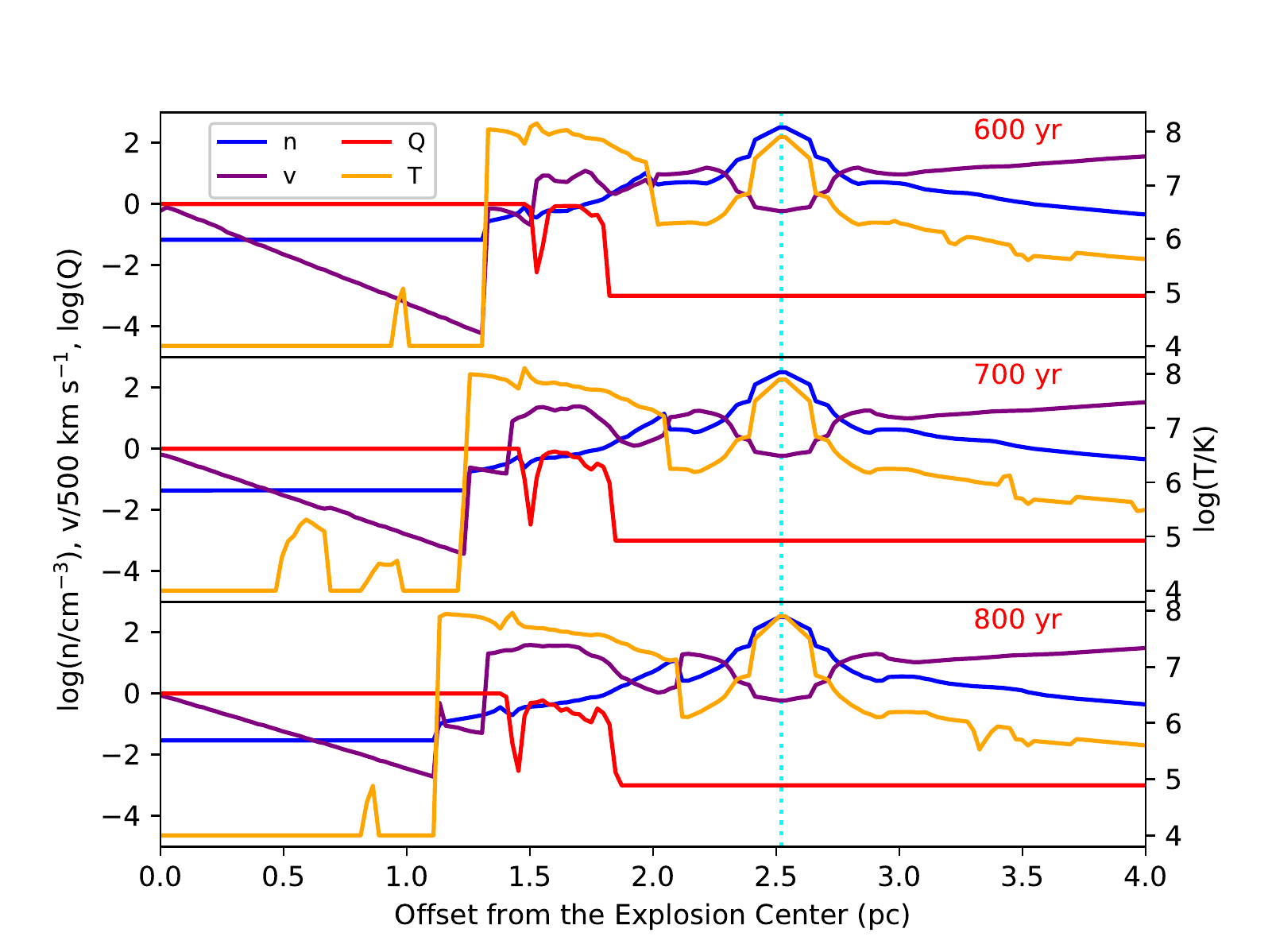}
\caption{One-dimensional profiles of density (blue), temperature (orange), velocity with respect to Sgr A* (purple; positive as away from Sgr A*) and ejecta tracer (red) along the line connecting the explosion center (zero-point) and Sgr A* (marked by a vertical dotted line) at $t =$ 600, 700 and 800 yr in the simulation run \textit{E5M1.3}. Here the tracer parameter assumes a lower bound of 0.001 for ease of visualization.  \label{fig:Hecut} }
\end{figure*}

The stratified structure of this interaction zone is highlighted in Figure~\ref{fig:Hecut}, in which the one-dimensional profiles of gas density, temperature, velocity and ejecta tracer parameter between the explosion center and Sgr A* are plotted for the epochs of $t = 600$, 700 and 800 yr.
At $t = 600$ yr, the two standing shocks are clearly captured by coincident temperature, density and velocity jumps, one separating the unshocked and shocked ejecta at an offset of $\sim$1.5 pc from the explosion center, and the other separating the unshocked and shocked outflow at an offset of $\sim$2.1 pc.
Both the shocked ejecta and the shocked outflow have their temperatures strongly raised to $\sim 10^7-10^8$ K.
The main contact discontinuity between the shocked ejecta and shocked outflow is located at an offset of $\sim$1.6 pc, but along this particular line the ejecta actually intrudes deeper due to the Rayleigh-Taylor instability, which helps to mix the shocked ejecta with the shocked outflow.
Nevertheless, the ejecta generally do not reach within $\sim$0.6 pc from Sgr A*.
At later times ($t =$ 700 and 800 yr), the inner standing shock gradually weakens and is pushed back toward the unshocked ejecta by the over-pressured shocked ejecta, whereas the outer standing shock roughly maintains its position.


The bottom two panels of Figure \ref{fig:He_tot_long} again show the synthetic X-ray surface brightness distribution at the different epochs.
Compared to Figure~\ref{fig:5}, the color scale here is adjusted such that the central outflow zone is suppressed to enhance visualization of the outer regions. A high surface brightness ridge (in yellow and orange color, with ${\rm log}S_{2-8} \gtrsim -14.7$) is formed as early as $t = 400$ yr, which arises from the freshly shocked outflow.
This ridge gradually moves toward Sgr A* until $t \approx 700$ yr, as the nuclear outflow is being pushed back by the SN forward shock.
Between $t= 600-800$ yr, both the location and morphology of this ridge roughly match the observed X-ray ridge (delineated by the red wedge), especially its lower (southeastern, between position angles $85^{\circ}-175^{\circ}$) and brighter half (Figure~\ref{fig:obs}).
However, the upper (northeastern, between position angles $5^{\circ}-85^{\circ}$) half of the observed X-ray ridge, which is about half as bright as the lower half, is not reproduced in the synthetic X-ray map.
Moreover, the simulation-predicted mean surface brightness of the ridge at $t = 700$ yr is only
$\rm 2.8\times10^{-15}~erg~s^{-1}~cm^{-2}~arcsec^{-2}$, which is substantially lower than the observed value of $\rm 6.1\times10^{-15}~erg~s^{-1}~cm^{-2}~arcsec^{-2}$ for the part of the ridge between position angles $85^{\circ}-175^{\circ}$.
We address possible causes for this discrepancy in Section~\ref{subsec:ridge}.
From $t = 900$ yr and beyond, the X-ray surface brightness of the ridge rapidly fades away, as the result of the weakening interaction between the ejecta and the nuclear outflow.




\begin{figure*}
\includegraphics[width=\textwidth]{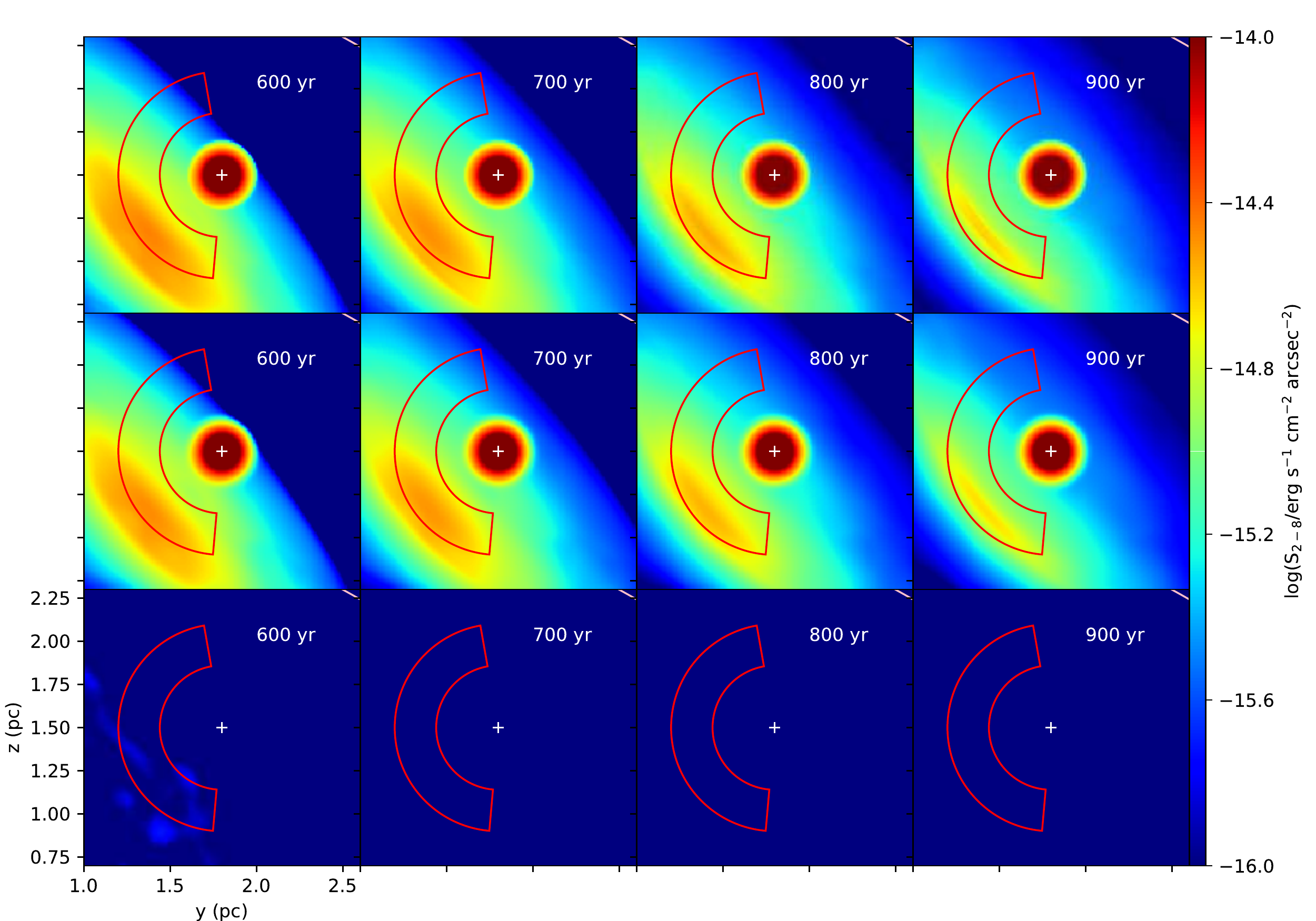}
\caption{Synthetic 2--8 keV X-ray surface brightness distribution of the 1.6 pc$\times$1.6 pc region centered on Sgr A*, smoothed with a 0\farcs5 Gaussian kernel and for snapshots from $t=$ 600 yr to 900 yr at a step of 100 yr, in the simulation run \textit{E5M1.3}. The top panels show the total surface brightness, while the middle and bottom panels show the contribution from the outflow-dominated and ejecta-dominated components, which are defined as having $Q < 0.001$ and $Q > 0.5$, respectively.
\label{fig:Heseq}}
\end{figure*}

Figure \ref{fig:Heseq} further displays the synthetic X-ray surface brightness maps between $t =$ 600 yr to 900 yr (top panel), this time smoothed with a Guassian kernel of FWHM = 0\farcs5 to mimic the {\it Chandra} point-spread function.
Meanwhile, we decompose the X-ray surface brightness into two components according to the tracer parameter $Q$: the outflow-dominated component with $Q < 0.001$ (middle panel) and the ejecta-dominated component with $Q > 0.5$ (bottom panel).
Obviously, it is the shocked outflow that dominates the bulk of the high-surface-brightness regions, in particular the ridge.
The shocked ejecta have a minor contribution to the surface brightness, mainly as clumps and filaments seen in the southeastern portion (most obvious at $t = 600$ yr), never reaching a surface brightness above $5\times10^{-16}\rm~erg~s^{-1}~cm^{-2}~arcsec^{-2}$.
These features, likely due to the Rayleigh-Taylor instability, gradually disappear after $t =$ 700 yr, consistent with the above notion that the ejecta is pushed back by the outflow.
That the shocked ejecta have a minor X-ray contribution in the ridge holds true even if a substantially higher (thus more realistic) metallicity were adopted to calculate the ejecta's X-ray emission (Appendix~\ref{app:metal}).


\subsection{The run with higher explosion energy}\label{subsec:he}

\begin{figure*}
\includegraphics[width=\textwidth]{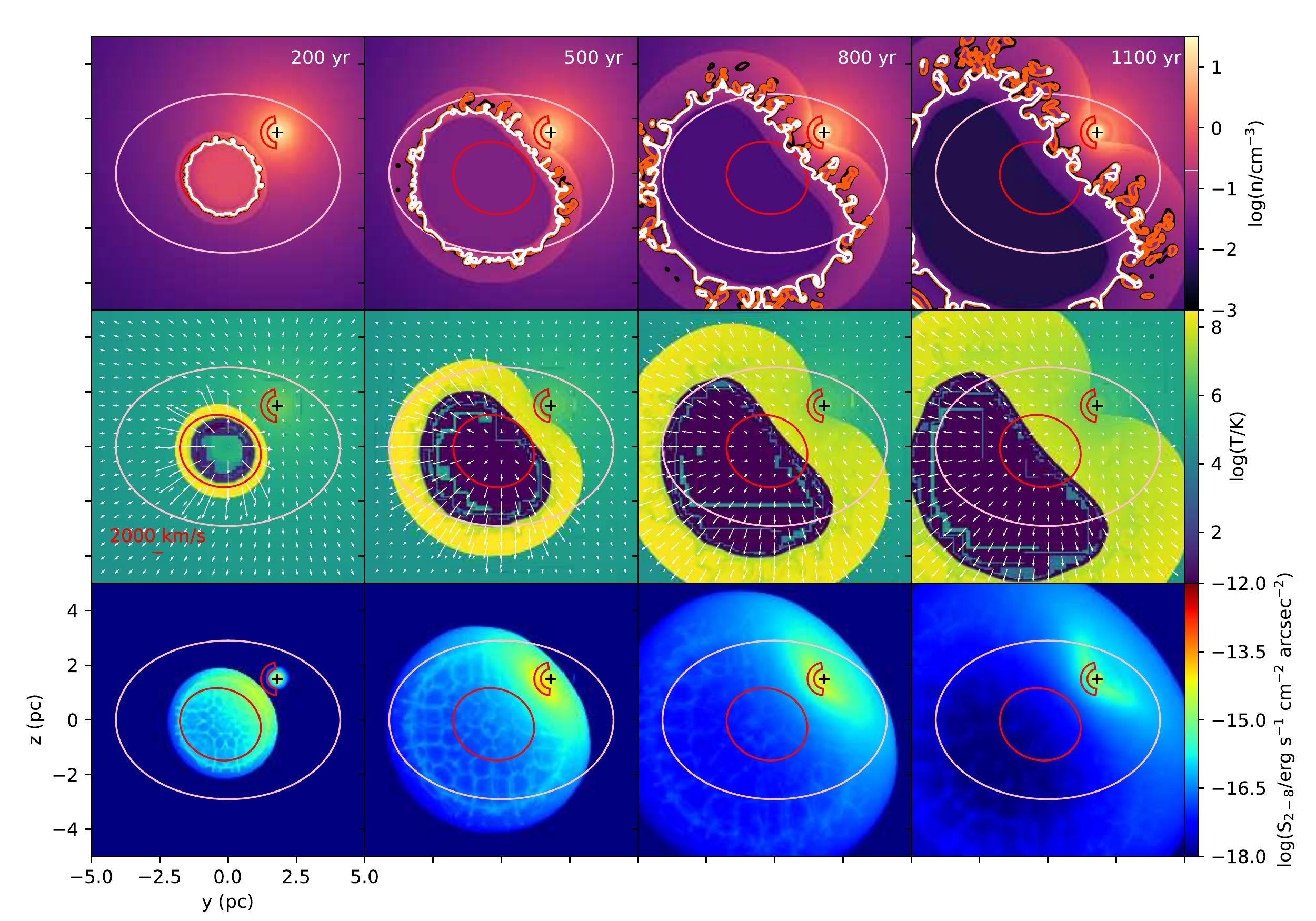}
\caption{Similar to Figure~\ref{fig:5}, but for simulation run \textit{E10M1.3} at $t=$ 200, 500, 800 and 1100 yr after the SN explosion.}
\label{fig:10}
\end{figure*}

\begin{figure*}
\includegraphics[width=\textwidth]{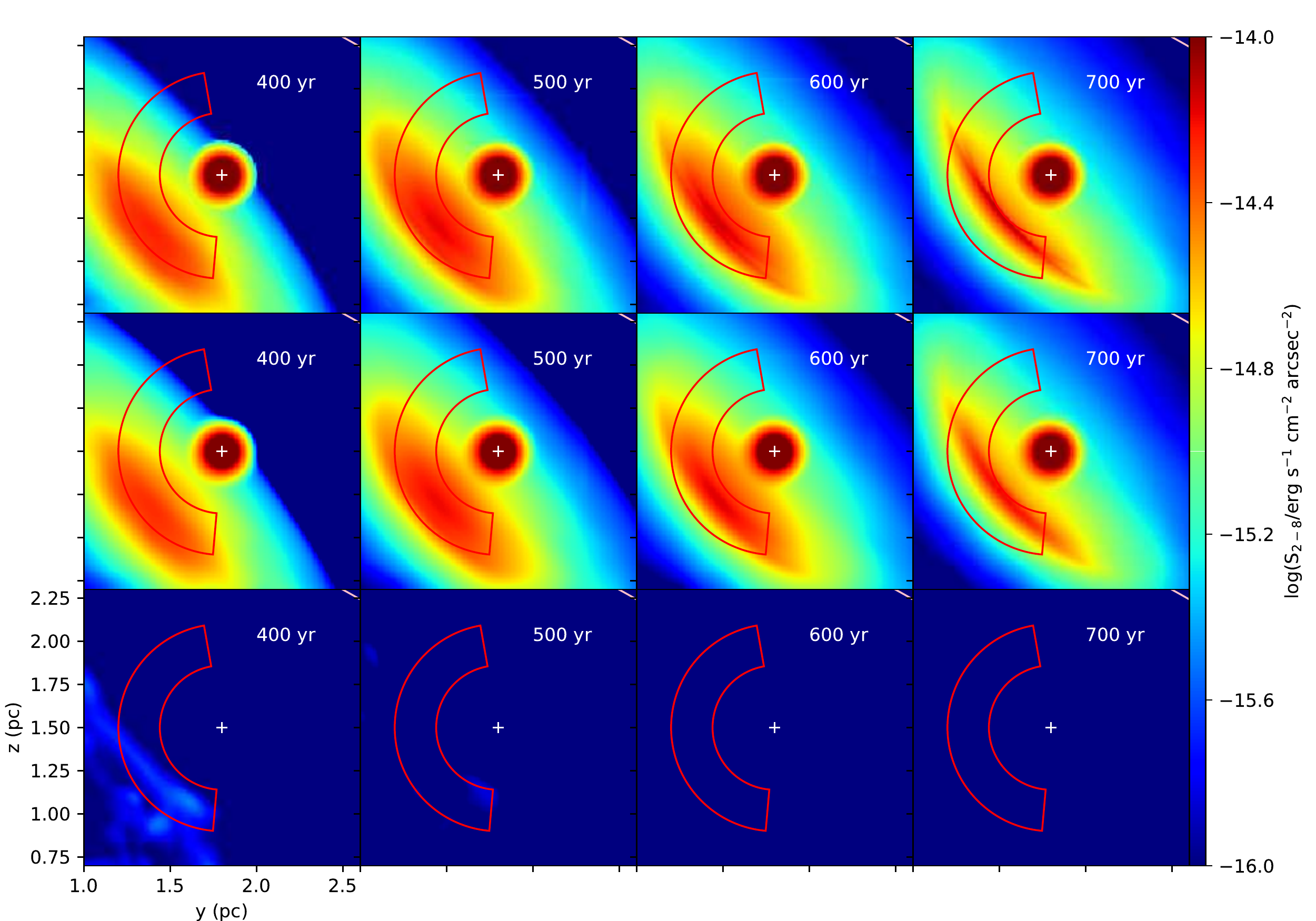}
\caption{Similar to the X-ray maps in Figure~\ref{fig:Heseq}, but for simulation run \textit{E10M1.3} at snapshots from $t=$ 400 yr to 700 yr at a step of 100 yr.
The maps have been smoothed with a 0\farcs5 Gaussian kernel.
The top panels show the total surface brightness, while the middle and bottom panels show the contribution from the outflow-dominated and ejecta-dominated components, which are defined as having $Q < 0.001$ and $Q > 0.5$, respectively.}
\label{fig:HeseqHE}
\end{figure*}

We now turn to simulation run \textit{E10M1.3}, which has two times higher explosion energy and the same ejecta mass compared to that assumed in the fiducial run, \textit{E5M1.3}.
Similar to Figure~\ref{fig:5},  Figure \ref{fig:10} displays snapshots of density, temperature and X-ray surface brightness at $t =$ 200, 500, 800 and 1100 yr in run \textit{E10M1.3}.
In this case, the overall evolution of the SNR is similar to the fiducial case.
Due to the higher explosion energy, the forward shock acquires a higher velocity, thus reaching the same physical extent at an earlier time than in the fiducial run.
In particular, except in the direction toward Sgr A*, the forward shock reaches an extent comparable to the observed size of the radio shell as early as $t = 500$ yr.
In the meantime, on its side approaching Sgr A*, the forward shock is interacting with the nuclear outflow, forming again a ridge of dense, high-temperature gas with high X-ray surface brightness, which is illustrated by the closeup view in Figure \ref{fig:HeseqHE}, for epochs from $t = 400$ yr to 700 yr.
However, the location of this ridge comes significantly closer to Sgr A* than in run \textit{E5M1.3} (compare Figure \ref{fig:He_tot_long}), which can be understood as due to the stronger momentum of the ejecta in run \textit{E10M1.3}.

At $t =$ 500 yr, when the overall morphology of the synthetic ridge agrees well with the observed one, we estimate a mean surface brightness of
$\rm4.9\times10^{-15}~erg~s^{-1}~cm^{-2}~arcsec^{-2}$.
This is higher than \textit{E5M1.3}, which is mainly due to the fact that the nuclear outflow has a higher density at a smaller distance from Sgr A* and that the X-ray emissivity is proportional to density squared.
However, this is still lower than the observed value and rapidly drops after $t =$ 600 yr.

The middle and bottom panels of Figure \ref{fig:HeseqHE} indicate that the overall X-ray emission in this region is also dominated by the shocked outflow.
The shocked ejecta contribute little to the X-ray emission from the ridge, although they reach a somewhat higher surface brightness in the southeastern portion of the map, when compared to the case of \textit{E5M1.3} (Figure~\ref{fig:Heseq}), likely due to the higher ejecta velocity that leads to a higher post-shock temperature.

\subsection{The run with higher ejecta mass}\label{subsec:cc}

\begin{figure*}
\includegraphics[width=\textwidth]{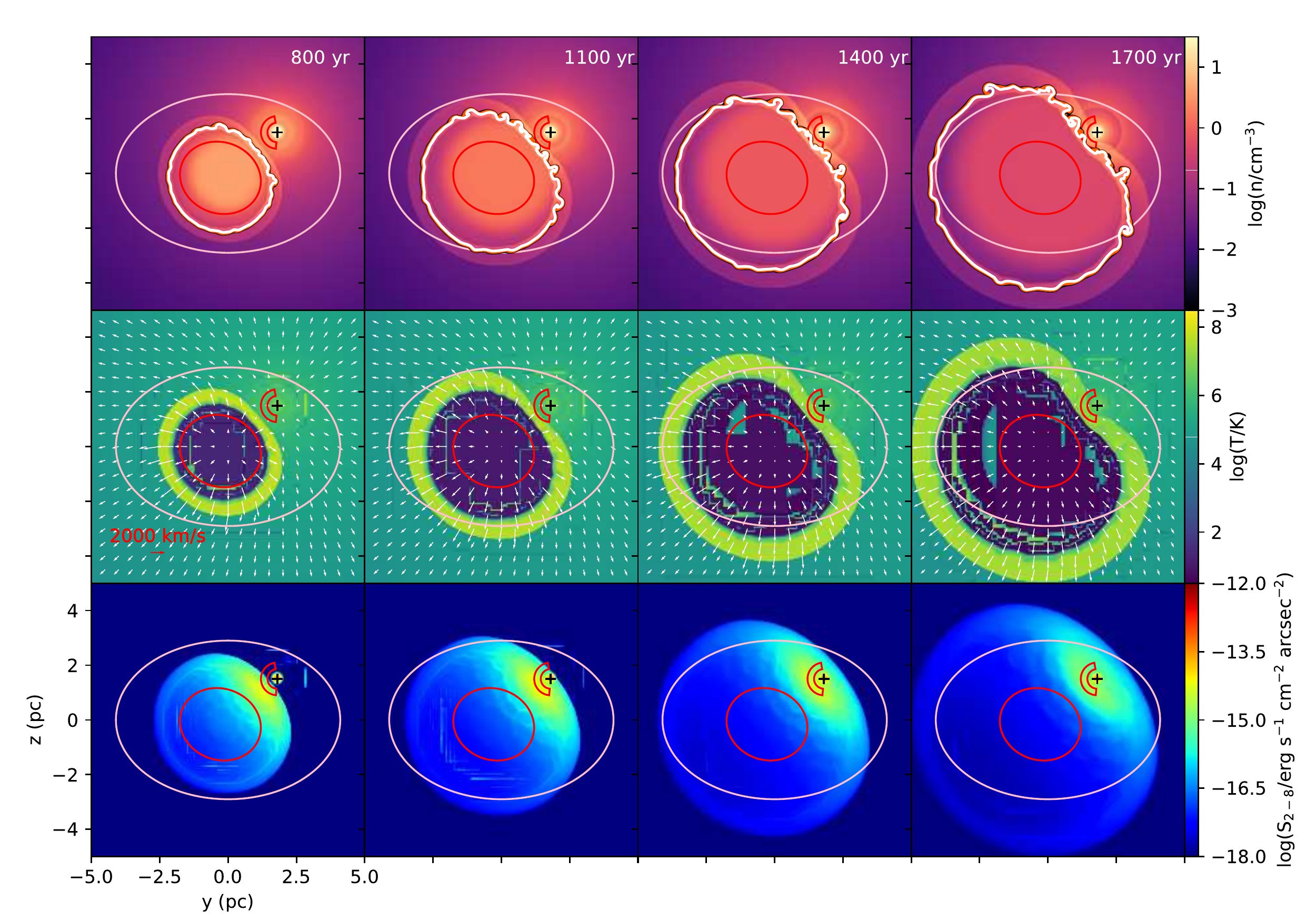}
\caption{Similar to Figure~\ref{fig:5}, but for simulation run \textit{E5M10} at $t =$ 800, 1100, 1400 and 1700 yr after the SN explosion.
\label{fig:He_tot_CC}}
\end{figure*}

\begin{figure*}
\includegraphics[width=\textwidth]{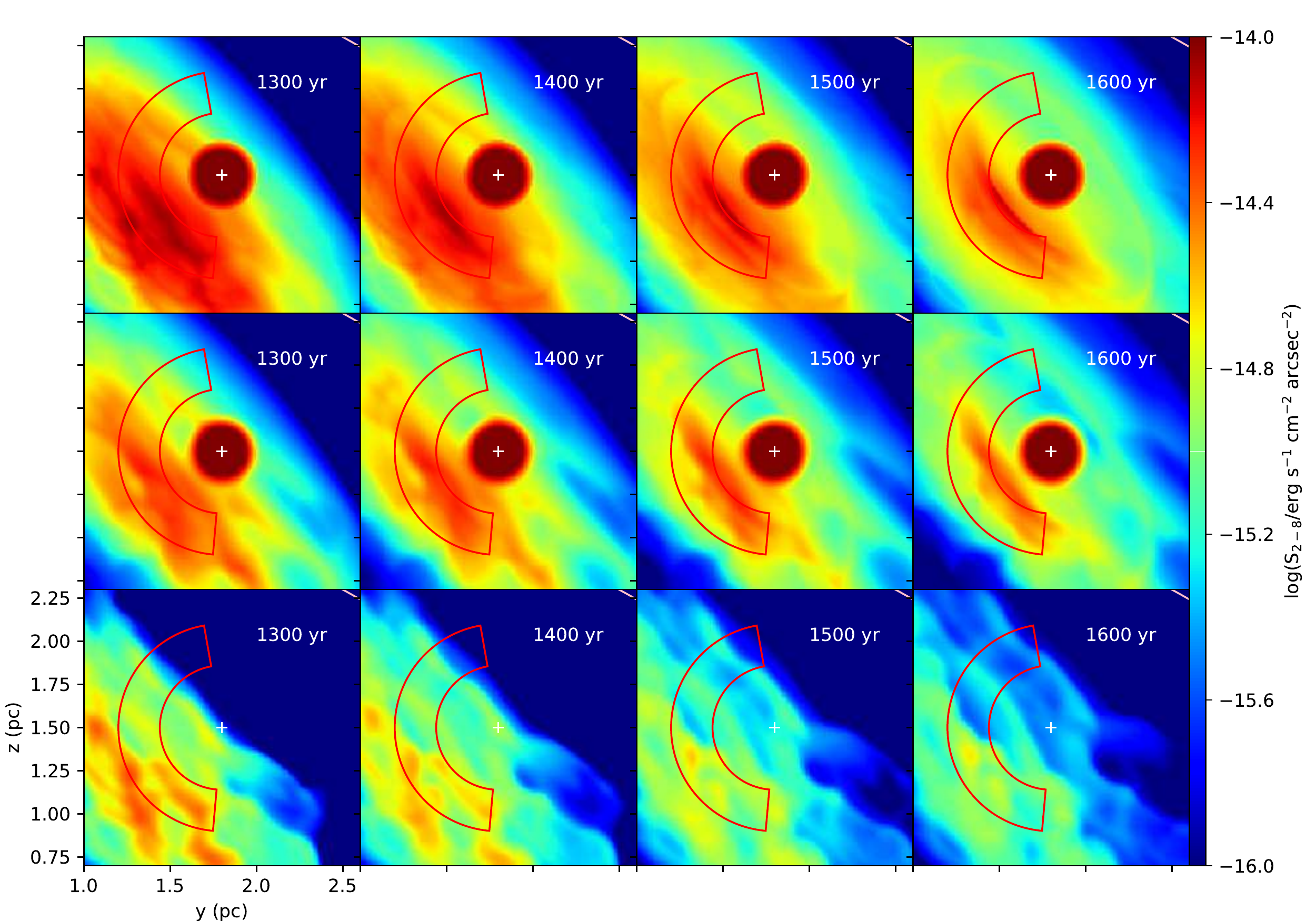}
\caption{Similar to the X-ray maps in Figure~\ref{fig:Heseq}, but for simulation run \textit{E5M10} at snapshots from $t=$ 1300 yr to 1600 yr at a step of 100 yr.
The maps have been smoothed with a 0\farcs5 Gaussian kernel.
The top panels show the total surface brightness, while the middle and bottom panels show the contribution from the outflow-dominated and ejecta-dominated components, which are defined as having $Q < 0.001$ and $Q > 0.5$, respectively.}
\label{fig:HeseqCC}
\end{figure*}

The simulation run \textit{E5M10} has the same explosion energy but a $\sim$8 times higher ejecta mass compared to that assumed in the fiducial run.
Figure \ref{fig:He_tot_CC} displays snapshots of density, temperature and X-ray surface brightness in run \textit{E5M10} at epochs of 800, 1100, 1400 and 1700 yr, which show that the overall evolution of the SNR is also similar to the fiducial case.
However, the higher ejecta mass, without an increase in the total energy, results in a substantially lower velocity of the forward shock and the ejecta.
Consequently, the SNR expands much slower, and not until $t =$ 1400 yr that the forward shock expands to a size roughly matching that of the radio shell.
In the meantime, the unshocked ejecta form a denser core compared to the fiducial case, again manifesting themselves as a cavity of little X-ray emission.
Similar to the two previous runs, a ridge of high X-ray surface brightness is formed due to interaction between the forward shock and the nuclear outflow, which is illustrated by the closeup view in Figure \ref{fig:HeseqCC} for epochs between $t =$ 1300 yr and 1600 yr.
The location of the ridge is closer to Sgr A* than in the fiducial run (Figure~\ref{fig:Heseq}), which can also be understood as due to a stronger momentum of the ejecta. Notably, the northeastern half of the observed ridge is again not reproduced.

At $t =$ 1400 yr, when the overall morphology of the synthetic ridge well matches the observed one, we estimate a mean X-ray surface brightness of the ridge as $\rm6.0\times10^{-15}~erg~s^{-1}~cm^{-2}~arcsec^{-2}$, which is quite comparable to the observed value.
That the mean X-ray surface brightness is substantially higher than in the fiducial case can be explained by a denser and slower ejecta (because of a higher $M_{\rm SN}$), which results in a higher thermal pressure in the shocked ejecta holding up a denser shocked outflow.

The middle and bottom panels of Figure~\ref{fig:HeseqCC} again compare the relative contributions of the shocked outflow and the shocked ejecta.
Notably, the shocked ejecta has a significantly higher fractional contribution at the apparent location of the ridge compared to the cases of \textit{E5M1.3} and \textit{E10M1.3}.
More quantitatively, the ejecta contribute $\sim$ 14\% of the X-ray emission from the ridge region.
However, by inspecting the 3D distribution of the tracer parameter, we find that the ejecta emission mostly comes from behind the ridge.
In the meantime, the shocked ejecta also produces significant emission at the southeastern corner, i.e., immediately outside the ridge. This appears coincident with the observed excess emission at a similar position, as evident in the right panel of Figure~\ref{fig:obs}.

\section{Discussion and Conclusion}\label{sec:dis}

In the previous sections, we have presented 3D simulations of the early-stage hydrodynamic evolution of a SNR in the central parsecs of the Galactic center, incorporating necessarily simplified yet sufficiently realistic physical conditions of both the SN explosion  and the environment in which it evolves.
The three main simulation runs, representing the evolution of a SN Iax (\textit{E5M1.3}), a SN Ia (\textit{E10M1.3}) or a CCSN (\textit{E5M10}), have a varied degree of success and failure in reproducing the main observed properties of Sgr A East (Section~\ref{sec:obs}).
In this section, we discuss the results and implications for our understanding of the enigmatic Galactic center ecosystem.

\subsection{Formation of the X-ray ridge and the progenitor supernova of Sgr A East}
\label{subsec:ridge}
\begin{figure}
\includegraphics[width=0.5\textwidth]{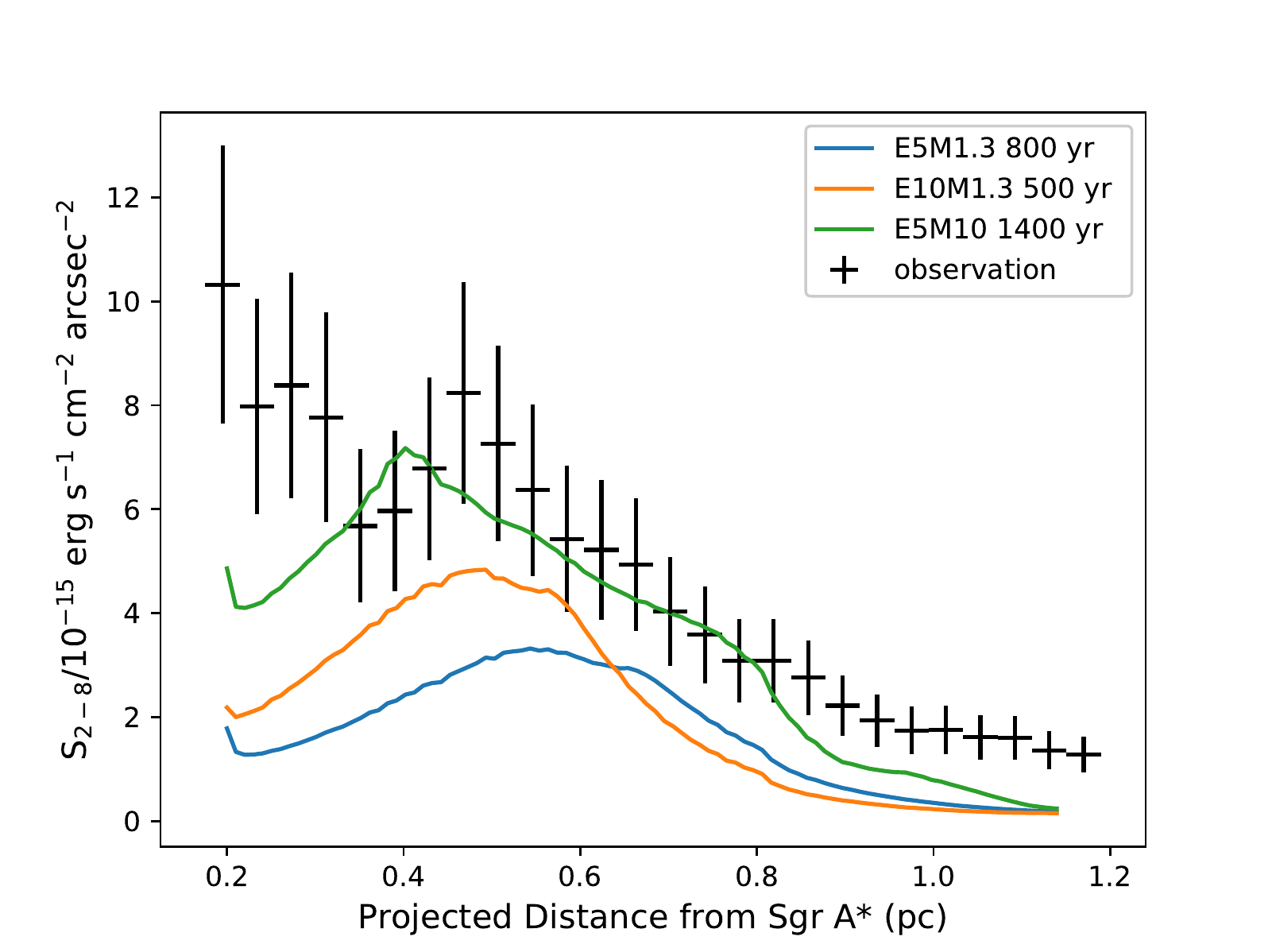}
\caption{Observed (black data points with error bars) and synthetic (colored curves) 2--8 keV surface brightness profiles, constructed from consecutive wedges centered on Sgr A* and covering position angles between $85^\circ-175^\circ$, at a certain epoch indicated by the insert.}
\label{fig:sb}
\end{figure}

All three simulations can reproduce a ridge-like structure, which is unambiguously identified as the site of dynamic interaction between the nuclear outflow and the expanding SNR, thus confirming the proposal of \citet{Rockefeller2005}.
We further demonstrate that the X-ray emission from the ridge is dominated by the shocked outflow and that the SN ejecta can barely penetrate into the ridge.
Among the three simulations, \textit{E5M10} produces a ridge a bit too close to Sgr A*, whereas in \textit{E5M1.3} the ridge finds itself at a location somewhat further away from Sgr A* compared with the observation (Figures~\ref{fig:Heseq} and \ref{fig:HeseqCC}). This is more concisely illustrated in Figure \ref{fig:sb}, which compares the X-ray radial surface brightness profiles of the three simulations, constructed from consecutive wedges centered on Sgr A* and covering position angles between $85^\circ-175^\circ$, for the epoch when the ridge has the best observation-matching morphology.
These synthetic profiles are to be contrasted with the observed one, which has been extracted from the deep {\it Chandra} image (Figure~\ref{fig:obs}) in the same manner, with additional care in background subtraction and masking of interloping point sources (Z. Hua et al. 2023, submitted).
Comparing the three profiles, a remarkable agreement between the observed and synthetic X-ray surface brightness of the ridge is clearly found in \textit{E5M10}, except that the peak of its profile is found at $\sim$0.4 pc from Sgr A*, which is $\sim0.08$ pc closer than actually observed.
In contrast, the profiles of \textit{E5M1.3} and \textit{E10M1.3} are substantially lower and flatter than observed.

The above comparison between the synthetic and observed properties of the X-ray ridge thus favors a CCSN as the progenitor supernova of Sgr A East.
This is in line with the suggestion or assumption of various previous work \citep[e.g.,][]{2002ApJ...570..671M,2004MNRAS.350..129S,Park2005,2013ApJ...777..146Z,Lau2015}, but
in tension with the recent work of \citet{2021ApJ...908...31Z}, which, based on X-ray spectroscopic evidence of a low ratio of intermediate-mass elements to Fe and large Mn/Fe and Ni/Fe ratios in the X-ray-emitting ejecta of Sgr A East, disfavored a CCSN but favored a SN Iax origin.

From the simulation point of view, there are potential ways to alter the synthetic X-ray emission from the ridge for a given type of SN explosion.
The X-ray surface brightness of the ridge is strongly affected by both $E_{\rm SN}$ and $M_{\rm SN}$.
 Current models for SN Iax \citep{2013ApJ...767...57F} leave little room for changing $M_{\rm SN}$ (i.e., not significantly below the Chandrasekhar limit).  Meanwhile our assumption of $E_{\rm SN} = 5~\times10^{50}\rm~erg$ is probably already close to the upper end for these under-luminous explosions.
Properties of the nuclear outflow are equally important in determining the X-ray surface brightness, as it is demonstrated in Section~\ref{sec:res} that X-ray emission from the ridge is dominated by the shocked outflow.
Our three simulations share a default nuclear outflow, for which we have adopted basic parameters consistent with the dedicated numerical simulations that establish the hydrodynamic behavior of the nuclear outflow \citep{2018MNRAS.478.3544R, 2020ApJ...888L...2C}.
In particular, a total mass loss rate of $\dot{M}_{\rm out} = 1\times10^{-3}\rm~M_{\odot}~yr^{-1}$ is assumed, which is inferred from near-infrared spectroscopic observations of the $\sim$30 individual WR stars \citep{Martins2007}.
A higher $\dot{M}_{\rm out}$, for instance $3\times10^{-3}\rm~M_{\odot}~yr^{-1}$ as suggested by older observations \citep{1986MNRAS.222..299G} and adopted by \citet{Rockefeller2005}, is expected to produce a substantially higher X-ray flux from the ridge. However, a higher $\dot{M}_{\rm out}$ also means a stronger ram pressure of the nuclear outflow, and consequently a much larger offset of the ridge from Sgr A* \citep{Yalinewich2017}. We have performed a test simulation with $\dot{M}_{\rm out} = 3\times10^{-3}\rm~M_{\odot}~yr^{-1}$, with other settings identical to \textit{E5M1.3}, to confirm this expectation.
Other default parameters, such as the chemical composition of the outflow and the ejecta, or the assumed CIE emissivity, are not expected to significantly alter the synthetic X-ray emission.
Therefore, it appears difficult to reconcile the $\sim$2 times lower X-ray surface brightness in \textit{E5M1.3}, without sacrificing the agreement found in the ridge location and the consensus on the WR star wind properties by previous observational and theoretical work.
The same above arguments can in fact be applied to the \textit{E5M10} case, suggesting that the putative CCSN has only a moderate explosion energy of $5\times10^{50}{\rm~erg}$. Raising this to a more canonical energy of $10^{51}{\rm~erg}$, for instance, would come at the price of a higher nuclear outflow rate, in order to reconcile with the ridge location (see Appendix \ref{app:std}).

Nevertheless, it is worth emphasizing that the spectroscopic evidence for a SN Iax origin as suggested by \citet{2021ApJ...908...31Z} is totally independent of the hydrodynamical arguments presented here. Further, the remarkable match between the \textit{E5M10} simulation and the observations of the morphology and X-ray flux of the ridge should be considered at least a partial coincidence, in view of the uncertainty in the assumed explosion energy, ejecta mass and outflow rate, as well as the uncertainty in the line-of-sight geometry.
Therefore, a final word on the progenitor SN must await further studies. For instance, the ejecta of a SN Iax may have a higher degree of spatial uniformity than that of a CCSN, which can be examined by a spatially-resolved X-ray spectral analysis \citep{2011ApJ...732..114L}.

\begin{figure}
\includegraphics[width=0.5\textwidth]{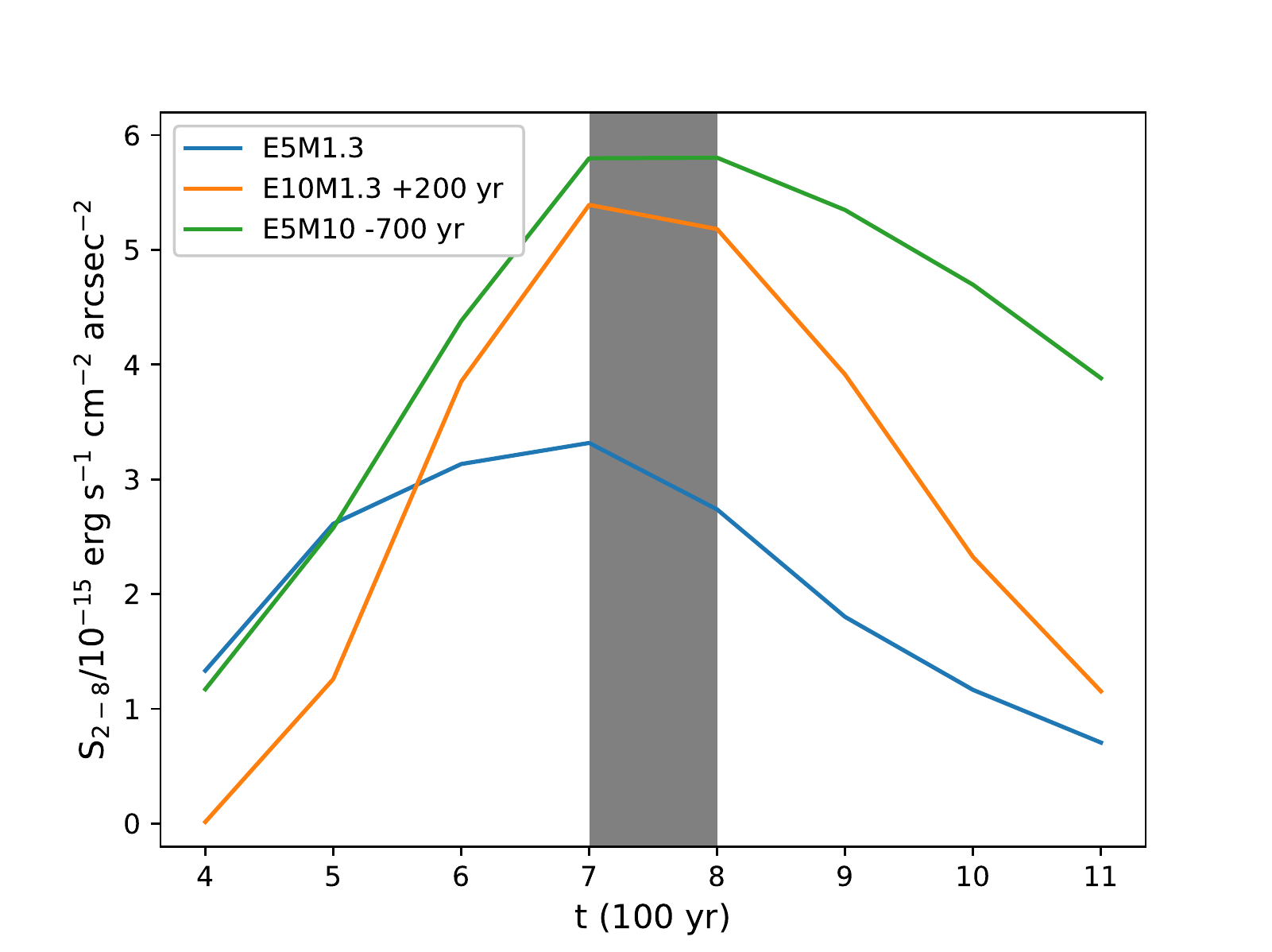}
\caption{The mean 2--8 keV surface brightness of the ridge as a function of time. The light curves of \textit{E10M1.3} and \textit{E5M10} have been artificially shifted for better alignment. A shadow between 700-800 yr is used to highlight the peak and a subsequent mild decay.}
\label{fig:curve}
\end{figure}

The future evolution of the ridge might provide an alternative diagnostic. As shown in Section~\ref{sec:res} and concisely illustrated in Figure~\ref{fig:curve},
the mean surface brightness of the ridge is predicted to decline shortly after the maximum value is reached in all three simulations.
From the synthetic light curves in Figure~\ref{fig:curve}, we estimate a fractional decay of 20\% and 4\% over the next 100 yr, followed by rapid decay afterwarsds, for the case of \textit{E5M1.3} and \textit{E10M1.3}, respectively. In the case of \textit{E5M10} the mean surface brightness is predicted to remain nearly constant for 100 yr, followed by a much slower decay afterwards.
Observationally, it is found that the X-ray surface brightness of the ridge exhibited no significant variability over the past 20 years as seen by the {\it Chandra} observations (Z. Hua et al. 2023, submitted), which further supports the case of \textit{E5M10}.
Future monitoring observations of the ridge into the next decades will provide a more definite answer.
In a few hundred years, the nuclear outflow is expected to completely resume its dynamical dominance in the central parsec, whereas the shocked ejecta will be pushed away and, on a longer timescale, cool and fragment until fully mixed with the circumnuclear medium.

We shall remark on the rather puzzling aspect that all three simulations fail to reproduce the northeastern half (position angles between $5^\circ$--$85^\circ$) of the X-ray ridge (Section~\ref{subsec:fidu}).
We note that this northeastern half is unlikely an interloping foreground/background feature, in view of its morphology that smoothly joins with the northwestern half of the ridge.
The most obvious way to generate an azimuthally-wider ridge is to have the SN ejecta coming from smaller position angles, but our test simulations show that this requires that the explosion center be placed 1 pc north of the currently assumed position  (see more discussions in Appendix \ref{app:pos}). Consequently, the position of the forward shock would look very different from that of the radio shell (Figure~\ref{fig:ls}). Moreover, due to the symmetry of the problem, a shifted explosion center cannot automatically lead to a factor of $\sim$2 difference in the observed X-ray surface brightness between the northeastern and southeastern halves of the ridge (Figure~\ref{fig:obs}).
We conclude that other factors are responsible for the wider ridge.
One such factor may be an anisotropic nuclear outflow. For ease of implementation we have assumed an isotropic, radial outflow, having azimuthally-averaged thermodynamic properties derived from the dedicated 3D simulations of wind-wind interactions within the central half-parsec region \citep{2018MNRAS.478.3544R}.
However, in reality the WR stars have a non-uniform and anisotropic distribution (Figure~\ref{fig:obs}), and most of them orbit about Sgr A* counterclockwise \citep{Genzel2010} in a period of $\sim10^4$ years.
These might have temporarily resulted in a weaker outflow at position angles between $5^\circ$--$85^\circ$, allowing the SN ejecta to reach a closer distance from Sgr A* than in the isotropic case.
 A collimated outflow driven by Sgr A* may cause additional anisotropy (e.g., \citealp{2013ApJ...779..154L}).
A similar argument might also be true for the SN ejecta, such that it temporarily carries a stronger momentum and intrudes to a closer distance along the desired orientation.
Alternatively and/or simultaneously, a significant angular momentum carried by the nuclear outflow, which is inherited from the counterclockwise orbital motion of the WR stars, might facilitate a laminar motion along the ridge, effectively increasing its azimuthal extent as observed.
Confirming these possibilities would require dedicated simulations incorporating the 3D distribution and motion of the WR stars, which we defer to future work.

We note in passing that G359.95-0.04, an X-ray-bright, comet-like, non-thermal feature (Figure~\ref{fig:obs}), which has long been considered a pulsar wind nebula \citep{2006MNRAS.367..937W}, may find its physical connection to the ridge.
Recently,  \citet{2022ApJ...927L...6Z} detected the long-sought compact radio counterpart of G359.95-0.04, which exhibits a proper motion consistent with the latter's head-tail morphology. Interestingly, the relative position of G359.95-0.04 and the ridge suggests that the putative pulsar is probably running into the ridge.
In this case, the ridge as a dense structure may provide the necessary ram pressure to form the observed head-tail morphology of G359.95-0.04, which is otherwise hard to explain, as the direction of the proper motion is such that the putative pulsar travels along, rather than against, the nuclear outflow.
It would be interesting to study how the forward shock of a SN propagating in the central parsec might ``light up'' hidden pulsars, which are expected to be numerous in the Galactic center \citep{2012ApJ...753..108W}.

\subsection{The age of Sgr A East}
\label{subsec:age}
All three runs can match the observed average size of Sgr A East, but at a substantial different epoch after the SN explosion, due to the different explosion energy and ejecta mass assumed.
Remarkably, this is achieved approximately at the same time as the formation of the X-ray ridge, in particular in simulations {\textit E5M1.3} ($t =$ 700 yr) and {\textit E5M10} ($t =$ 1400 yr).
Notably from Figures~\ref{fig:5}, \ref{fig:10} and \ref{fig:He_tot_CC}, the eastern half of the shell, which is unaffected by the nuclear outflow, appears rounder than actually observed. This can be understood as due to the intrinsic symmetry assumed in both the SN injection and the ambient gas. In reality, SN ejecta are often found to exhibit a certain degree of anisotropy.
In addition, it is known that the forward shock of Sgr A East is running into and interacting with the so-called {\it 50 km s$^{-1}$ Cloud} to the southeast (\citealp{2008ApJ...674..247L, 2012A&A...540A..50F}; Figure~\ref{fig:obs}), which is at least partially responsible for the flattened morphology.
We note also that we have not attempted to calculate a synthetic radio synchrotron intensity for the shell, which requires knowledge about the local magnetic field and energy density of relativistic electrons accelerated by the forward shock.
We defer to future work a more dedicated simulation to self-consistently and quantitatively reproduce the radio shell.
Nevertheless, we consider that all three simulations are able to reproduce the radio shell of Sgr A East.
This alone constrains the age of Sgr A East between $\tau_{\rm SN}\sim$ 500 yr to $\sim$ 1500 yr, depending on whether the progenitor SN was a SN Ia or a CCSN. We note that the latter case is in agreement with \citet{Rockefeller2005}, who inferred an age of $\sim1700$ yr, based on SPH simulations of a CCSN interacting with the nuclear outflow, although their very different initial conditions compared to ours necessarily led to very different evolutionary paces.
More recently, \citet{2022A&A...668A.124E} also inferred an age of 1--2 kyr for Sgr A East, in their simulations with a low density medium, although the age can reach 10 kyr in their simulations with a dense medium near the explosion center.
However, they and we do not reproduce the ”X-ray ridge” in the case with an age of 10 kyr, so the age derived from our simulations is favoured on this point.

That $\tau_{\rm SN}\lesssim$ 1500 yr calls into serious question an often assumed physical association between Sgr A East and the Cannonball \citep{Park2005,2013ApJ...777..146Z}.
Given the current position of the Cannonball (Figure~\ref{fig:obs}) and its measured transverse velocity of $500\pm100\rm~km~s^{-1}$ \citep{2013ApJ...777..146Z}, a physical association with Sgr A East requires a dynamical age of at least 6400 yr, which is highly unlikely given the inferred $\tau_{\rm SN}$.
We note that a similar conclusion was first drawn by \citet{Yalinewich2017}, who used hydrodynamic simulations to study SNR evolution in the Galactic center under more general conditions.
This leaves the relic of the progenitor SN an open question, although the existence of a relic neutron star is neither expected (in the case of SN Ia or Iax) 
nor required (in the case of a CCSN giving birth to a black hole).
Even assuming that a relic neutron star acquired a transverse velocity similar to that of the Cannonball, which is already high among observed pulsar kicks \citep{2006ApJ...643..332F},
it would have travelled a projected distance of only 0.5($\tau_{\rm SN}$/1000 yr) pc, i.e., it should still be found deep inside the SNR.
This invites a target search, perhaps most beneficial among known compact X-ray sources \citep{2018ApJS..235...26Z} and radio sources in the region \citep{2020ApJ...905..173Z}.
On the other hand, our simulations predict that the forward shock has a current expansion velocity of $\sim2000\rm~km~s^{-1}$. Thus in the near future the SNR shell is expected to catch up with and encompasses the Cannonball, certainly in projection and plausibly in real space.
Interestingly enough in the latter case, the Cannonball would become the Cuckoo Egg.

\subsection{The unexplained X-ray-bright interior}
\label{subsec:disevo}

There is yet another important observational aspect that all three simulations fail to reproduce, namely, the X-ray-emitting interior of Sgr A East, based primarily on which elemental abundances of the SN ejecta were measured and the type of progenitor SN were debated \citep{2002ApJ...570..671M,2004MNRAS.350..129S,2021ApJ...908...31Z}.

As illustrated in Figures~\ref{fig:5}, \ref{fig:10} and \ref{fig:He_tot_CC}, the central region of the SNR remains cold (with temperatures $\lesssim10^4$ K), due to adiabatic expansion, and produces essentially no X-ray emission.
In general, the expanding SN ejecta can be heated up to X-ray-emitting temperatures ($\gtrsim 10^6$ K) by the reverse shock.
In previous work \citep{2002ApJ...570..671M,2004MNRAS.350..129S}, which took Sgr A East as an evolved SNR with an age of $\sim$5000--10000 yr, the reverse shock is supposed to have propagated back to the explosion center, heating the entire ejecta to X-ray-emitting temperatures.
This situation was also implicitly assumed by \citet{2021ApJ...908...31Z}.

However, in our simulations, the age of Sgr A East is constrained to be $\lesssim$1500 yr. In this case, the bulk of the ejecta are still unaffected by the reverse shock and remain cold, as illustrated in Figures~\ref{fig:5}, \ref{fig:10} and \ref{fig:He_tot_CC}.
A simple estimate following the model of \citet{Leahy2017a} suggests that an ambient density of at least $\sim 1\rm~cm^{-3}$ around the explosion center is required to have the reverse shock propagating back to the center in 2000 yr after the explosion.
However, in our simulations, the initial gas density in the vicinity of the explosion center is only $\rm \sim0.1~cm^{-3}$, which is essentially determined by the uninterrupted nuclear outflow (Figure~\ref{fig:init}).
The initial ambient gas density remains lower than $\rm 1~cm^{-3}$ even if the explosion center moves closer to Sgr A* until about 1 pc away, which is unlikely given the current shell position of Sgr A East.
In fact, in Figure~\ref{fig:sb} we already observe a deficiency of our predicted surface brightness outside the ridge (i.e., at a projected distance $\gtrsim$0.6 pc), which is understood to be dominated by the shocked ejecta.
This means that even at the site where the back reaction to the ejecta is strongest, the observed X-ray emission arising from the ejecta is largely unaccounted for.

A much higher ambient gas density than offered by the nuclear outflow can in principle help to reconcile with the observed X-ray-bright, metal-enriched ejecta of Sgr A East.
One possibility is that the SN exploded in a dense circumstellar wind-blown bubble, for which case a massive progenitor star (hence a CCSN) is most likely responsible.
Alternatively, the SN might have exploded inside the dense cloudlets of molecular gas.
Molecular gas is not rarely seen in the Galactic center.
However, the hot and fast nuclear outflow is expected to blow up or evaporate small cloudlets within the central parsecs. Thus the chance of a SN exploding inside some cloudlets should be rather slim.
Nevertheless, we show a test simulation in Appendix~\ref{app:CL} to illustrate the potential effect of dense cloudlets surrounding the SN explosion center, but caution that it is practically difficult to constrain the extra properties (e.g., density, size and number) of such cloudlets.
Moreover, the lopsided morphology of the X-ray interior, in the sense that the part facing the direction of Sgr A* is substantially brighter (Figure~\ref{fig:obs}), suggests some casual relation with the nuclear outflow or Sgr A* itself, whereas an initially high ambient density, whether due to the progenitor star or a dense molecular cloudlet, is not expected to introduce such a lopsidedness.
 In addition, the CND may partly impede and/or reflect the forward shock \citep{2016ApJ...817..171Z}.
A CND-like component was included in the simulation of \citet{Rockefeller2005}.
We have similarly performed a test simulation with the CND included, which is mimicked by many clumps of dense cold gas (with $n \sim 10^4\rm~cm^{-3}$ and $T \sim 100$ K) distributed along a circular orbit with a diameter of 3 pc and aligned with the Galactic disk \citep{2012A&A...540A..50F}.
We find that, consistent with \citet{Rockefeller2005}, the CND can only slightly affect the shock  propagation, but has no appreciable effect in either the formation of the ridge or the reverse shock-heating of the ejecta.

The current location of the reverse shock is also closely related to the detection of infrared-emitting dust within Sgr A East \citep{Lau2015}. It was suggested that the dust formed in the ejecta of a CCSN and survived the subsequent destruction caused by the passage of the reverse shock \citep{Lau2015}.
However, if Sgr A East were as young as our models imply, then the dust found deep inside the SNR -- if it indeed formed from the ejecta -- would not have encountered a reverse shock, according to our simulations. Therefore, one cannot argue using Sgr A East that dust can survive a reverse shock, which would call the conclusion of \citet{Lau2015} into question.

We speculate that the X-ray-bright ejecta are mainly attributed to photoionization rather than the conventional collisional ionization by the reverse shock.
Signatures of a photoionized and recombining plasma in the X-ray spectrum of Sgr A East were recently suggested by \citet{2019PASJ...71...52O}.
Such an intriguing situation can occur in the unique Galactic center environment, in the presence of a SMBH. While being currently extremely under-luminous, Sgr A* is known to exhibit a bolometric luminosity orders of magnitude higher just several hundred years ago \citep{2013ASSP...34..331P, 2013A&A...558A..32C}.
This activity coincides with the early evolutionary stage of Sgr A East, and to some extent would inevitably irradiate the expanding SN ejecta with copious hard X-ray photons.
If true, such a scenario can provide a natural understanding to the lopsided morphology of the X-ray-bright ejecta, despite the absent role of the reverse shock.

\section*{Acknowledgements}
This work is supported by the National Natural Science Foundation of China (grants 12225302, 11873028) and the National Key Research and Development Program of China (grant 2017YFA0402703).
We thank Dr. Ping Zhou and Fangzheng Shi for helpful discussions. We acknowledge the cosmology simulation database (CSD) in the National Basic Science Data Center (NBSDC) and its funds the NBSDC-DB-10.

\section*{Data Availability}
The simulation data underlying this article may be shared upon reasonable request to the corresponding author.

\bibliographystyle{mnras}
\bibliography{mydb}

\appendix

\section{The effect of changing the explosion center}
\label{app:pos}
As discussed in Section~\ref{sec:obs} and Section~\ref{subsec:config}, the exact position of the explosion center of Sgr A East is subject to some uncertainty, both on the sky plane and along the line-of-sight.
To explore the effect of a different position of
the explosion center (with respect to Sgr A*), we have performed several test observations.
Specifically, we adopt the following coordinates for Sgr A*: $(x, y, z)_{\rm BH}$ = (1.0, 1.8, 0.5), (1.0, 2.5, 1) and (0.5, 1.8, 1.5) pc. These are offset from the fiducial site, $(x, y, z)_{\rm BH}$ = (1.0, 1.8, 1.5). The remaining parameters of each run are identical to run {\it E5M1.3}.

\begin{figure*}
\begin{flushright}
\includegraphics[width=0.95\textwidth]{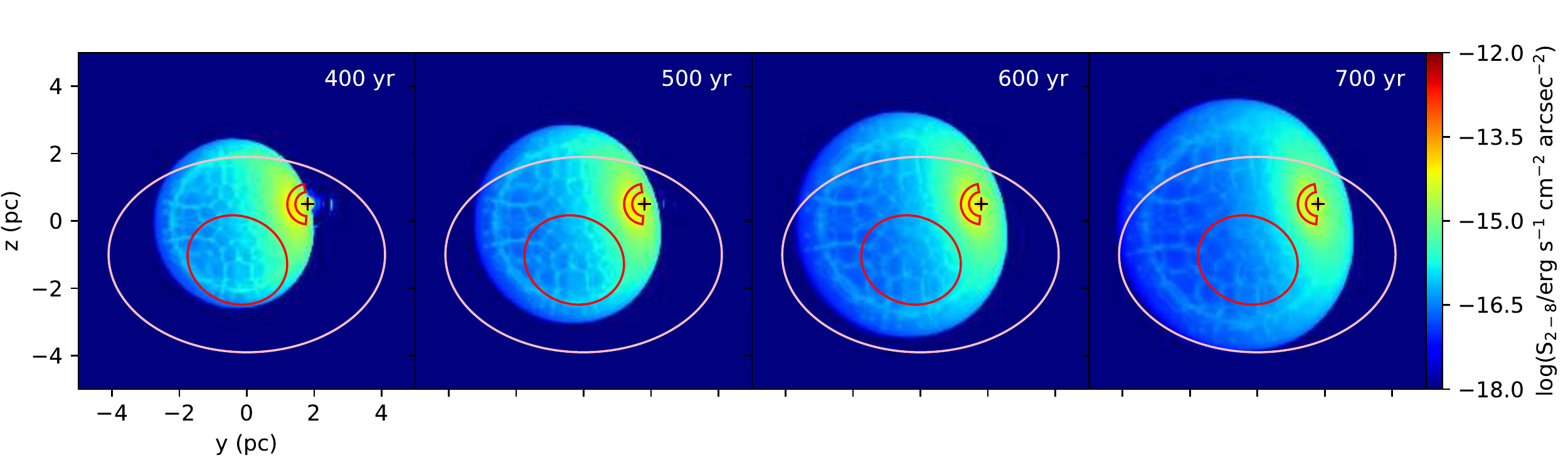}
\vskip-0.5cm
\includegraphics[width=0.98\textwidth]{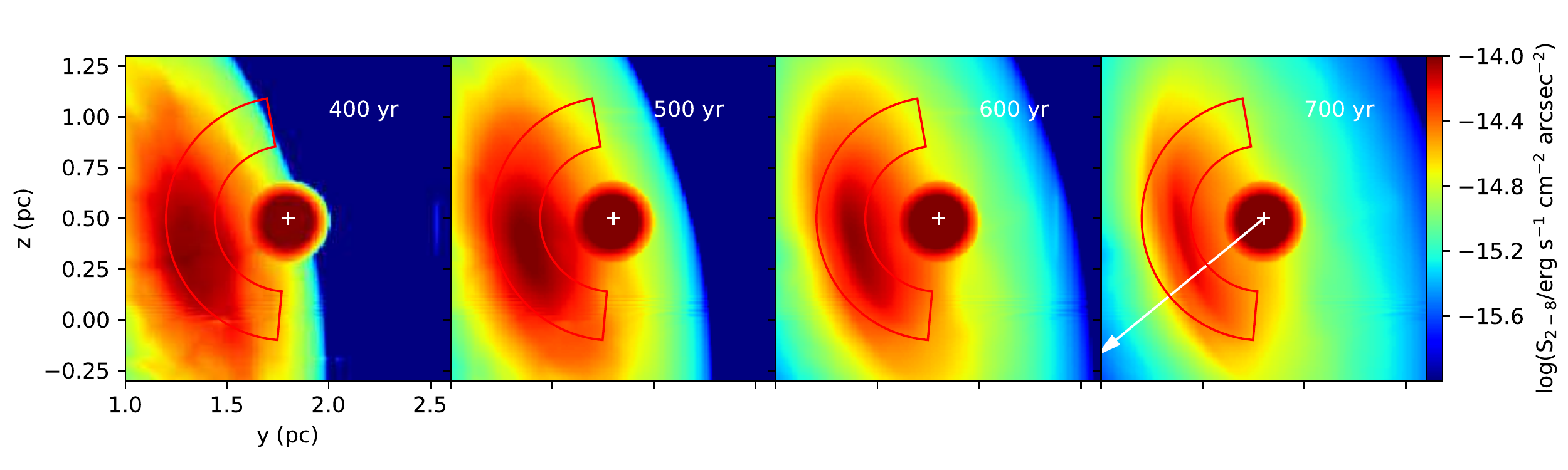}
\end{flushright}
\caption{The 2--8 keV X-ray surface brightness distribution of a test similuation similar to \textit{E5M1.3}, but with Sgr A* placed at $(x, y, z)_{BH} =(1.0, 1.8, 0.5)$ pc. The {\it top} panels show the snapshots in the whole simulation box at 400 -- 700 yr (from left to right) after the SN explosion,
while the {\it bottom} panels show the corresponding zoom-in images surrounding Sgr A*, which is marked by a black `+'.
The pink ellipse, red ellipse and red wedge outline the observed position of the radio shell, X-ray-bright interior and X-ray ridge, respectively, as in Figure~\ref{fig:obs}.
}.
\label{fig:ls}
\end{figure*}

As expected, the resultant morphology of the SNR and the X-ray ridge varies significantly with the assumed explosion center.
In particular, in the simulation with $(x, y, z)_{\rm BH} =(1.0, 1.8, 0.5)$ pc, i.e., the explosion center moves northward relative to Sgr A*, the predicted X-ray ridge can better reproduce the observed one, in particular the northeastern half, which is a notable discrepancy in the three main simulations (Section~\ref{subsec:ridge}).
This is illustrated in Figure~\ref{fig:ls}.
However, this improvement in reproducing the X-ray ridge comes at the price of
a substantially offset SNR, in accordance with the new explosion center, which cannot reproduce the observed location of the radio shell (nor for the location of the X-ray-bright core), as again illustrated in Figure~\ref{fig:ls}.
Therefore, it is unlikely that the true explosion site of Sgr A East is close to this northern position.

\begin{figure*}

\begin{flushright}
\includegraphics[width=0.95\textwidth]{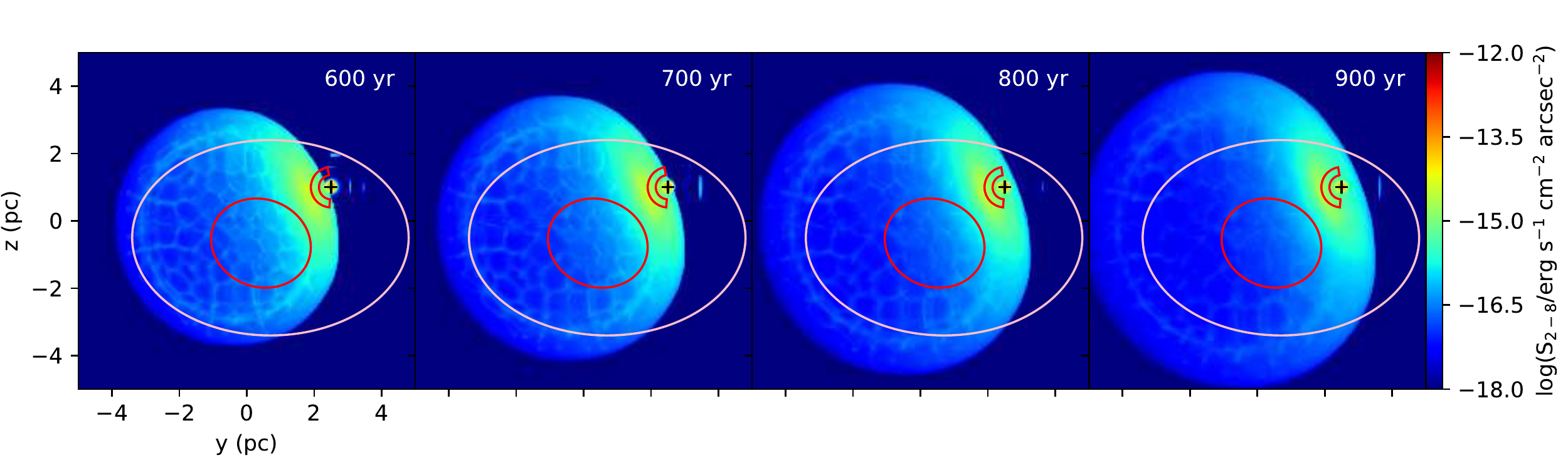}
\vskip-0.5cm
\includegraphics[width=0.98\textwidth]{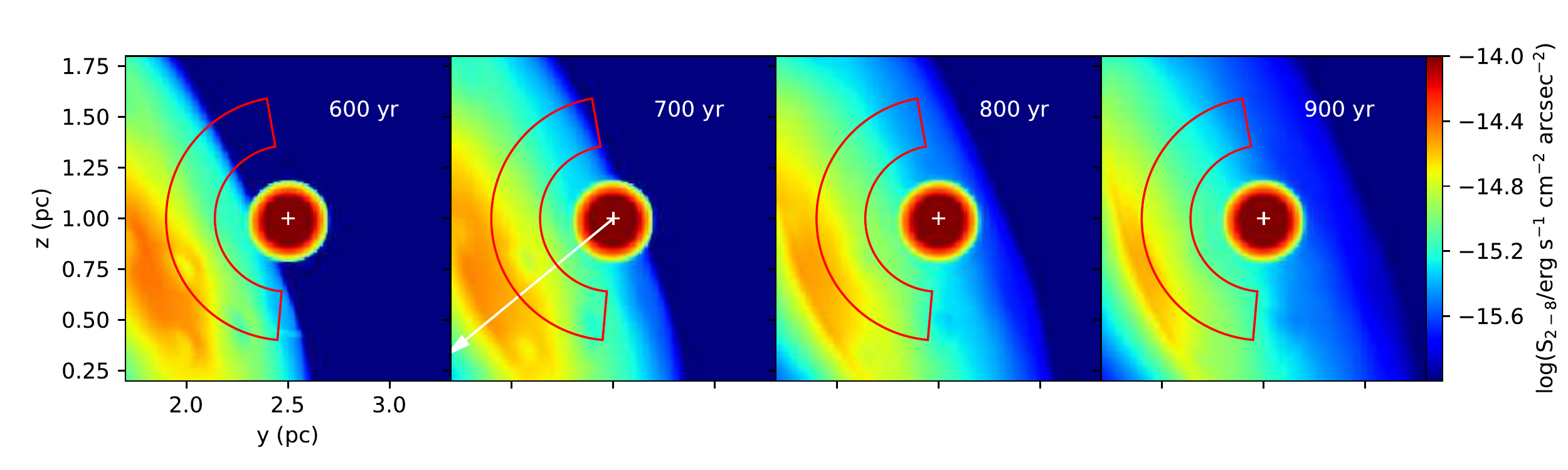}
\end{flushright}
\caption{Similar to Figure~\ref{fig:ls}, but with Sgr A* placed at (x,y,z)$_{\rm BH}$ = (1.0, 2.5, 1.0) pc. }.
\label{fig:12.51}
\end{figure*}

In the cases where the explosion center moves along both the $y$- and $z$-axis, i.e.,  $(x, y, z)_{BH}$ = (1.0, 2.5, 1) pc, the resultant morphology is clearly inconsistent with the observed, as illustrated in Figure~\ref{fig:12.51}.
In particular, the ridge appears much further away from Sgr A* in this case, due to a larger offset of the explosion center.

\begin{figure*}
\begin{flushright}
\includegraphics[width=0.95\textwidth]{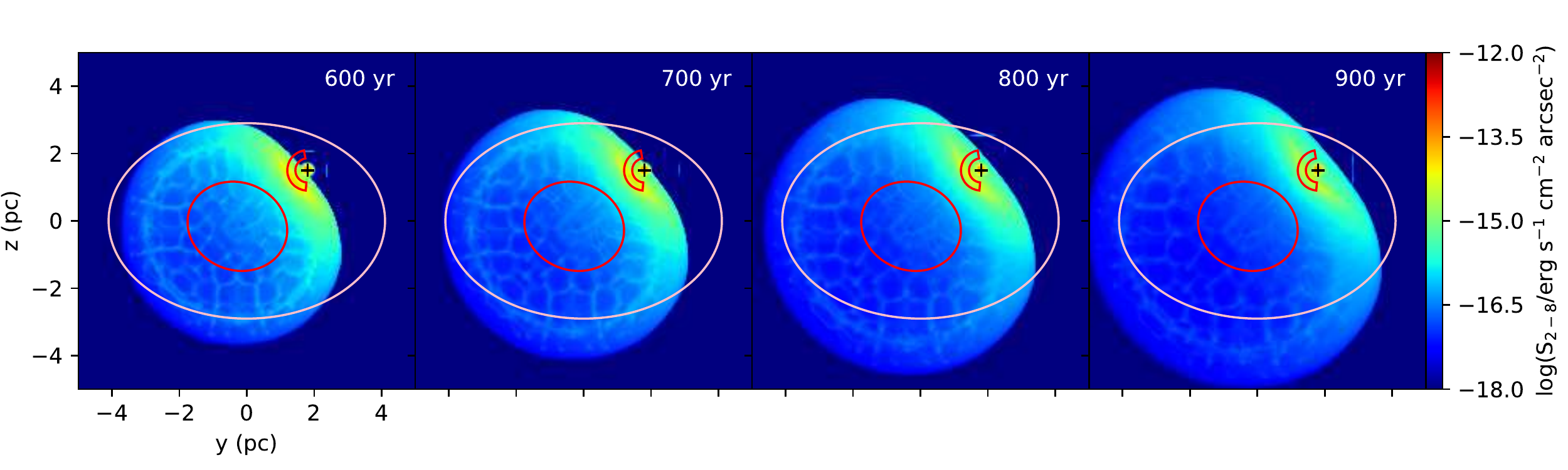}
\vskip-0.5cm
\includegraphics[width=0.98\textwidth]{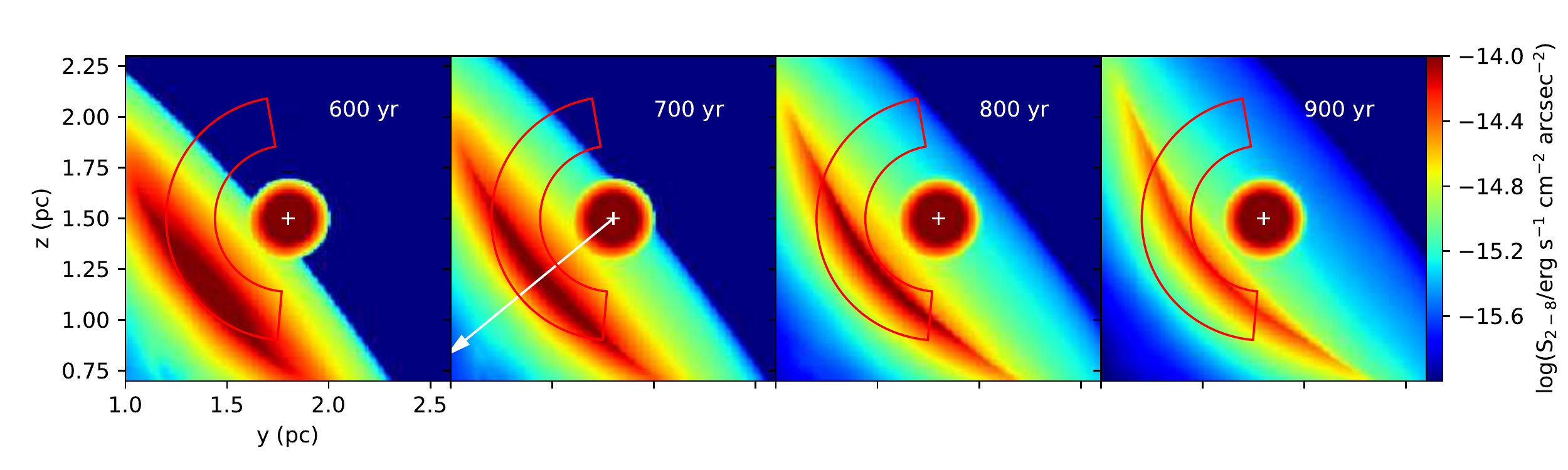}
\end{flushright}
\vskip-0.5cm
\caption{Similar to Figure~\ref{fig:ls}, but with Sgr A* placed at $(x,y,z)_{\rm BH} = (0.5, 1.8, 1.5)$ pc and for epochs of 600 -- 900 yr. }.
\label{fig:585}
\end{figure*}

Lastly, when the explosion center is placed further behind Sgr A*, i.e., $(x, y, z)_{BH}$ = (0.5, 1.8, 1.5) pc, the resultant ridge is still found close to the observed position, but the ridge appears thinner and brighter than observed (Figure~\ref{fig:585}), because in this case the interaction zone between the nuclear outflow and the SN shock projects into a narrower region.

\section{The case of a core-collapse supernova with a higher explosion energy}
\label{app:std}
\begin{figure*}
\includegraphics[width=\textwidth]{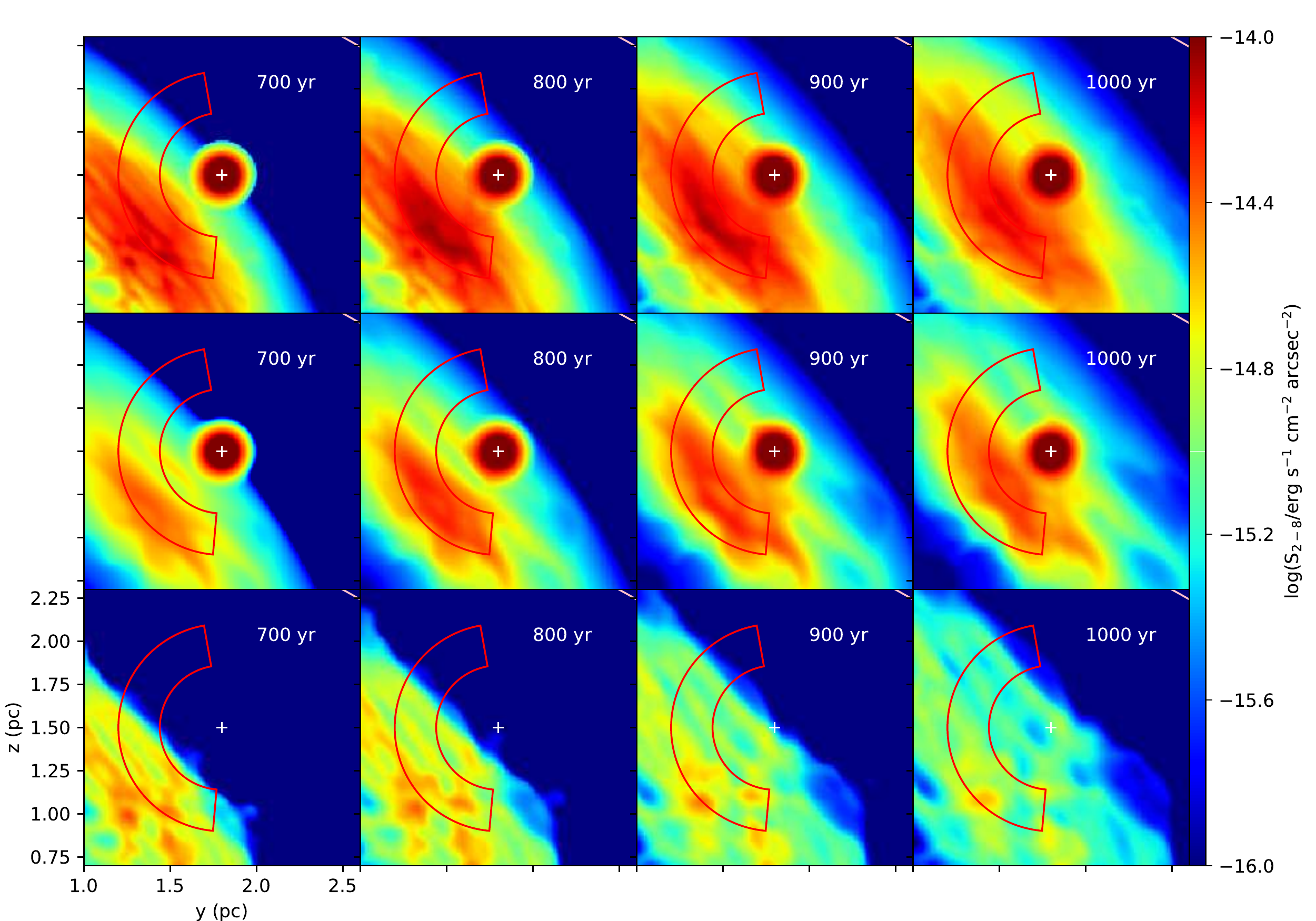}
\caption{Similar to the X-ray maps in Figure~\ref{fig:Heseq}, but for simulation run \textit{E10M10} at snapshots from $t=$ 700 yr to 1000 yr at a step of 100 yr. The maps have been smoothed with a 0\farcs5 Gaussian kernel.The top panels show the total surface brightness, while the middle and bottom panels show the contribution from the outflow-dominated and ejecta-dominated components, which are defined as having $Q < 0.001$ and $Q > 0.5$, respectively.}
\label{fig:HECC}
\end{figure*}

We have also performed the simulation \textit{E10M10}, which assumes an explosion energy of $10^{51}$ erg and an ejecta mass of 10 $\rm M_\odot$, with other settings identical to run ${\it E5M10}$.
However, in the present case the X-ray ridge forms earlier, due to the higher explosion energy (thus also a higher momentum).
As a result, the age of the system upon forming the ridge is significantly younger than in the case of \textit{E5M10}.
Figure~\ref{fig:HECC} shows the X-ray view of the ridge at several epochs in \textit{E10M10}.
In this case, the X-ray emission from the ridge is still dominated by the nuclear outflow, with only a small ($\sim$15\%) contribution from the ejecta at $t = 900$ yr (at this time the ridge looks most similar to the observed one).
A mean X-ray surface brightness of $\rm7.3\times10^{-15}~erg~s^{-1}~cm^{-2}~arcsec^{-2}$ is found for the ridge at $t =$ 900 yr, which
is somewhat higher than the observed value ($\rm6.1\times10^{-15}~erg~s^{-1}~cm^{-2}~arcsec^{-2}$).


\section{The effect of a higher metallicity in the SN ejecta}
\label{app:metal}
In the three main simulations we have assumed that the metallicity of the SN ejecta is the same as that of the nuclear outflow. In reality, the SN ejecta may have a substantially higher abundance for certain metals.
We have performed a test simulation in which a metallicity of six times solar is assumed for the SN ejecta, a value based on the iron abundance obtained by \citet{2021ApJ...908...31Z}.
The other settings are identical to run {\it E5M1.3}.

\begin{figure*}
\includegraphics[width=\textwidth]{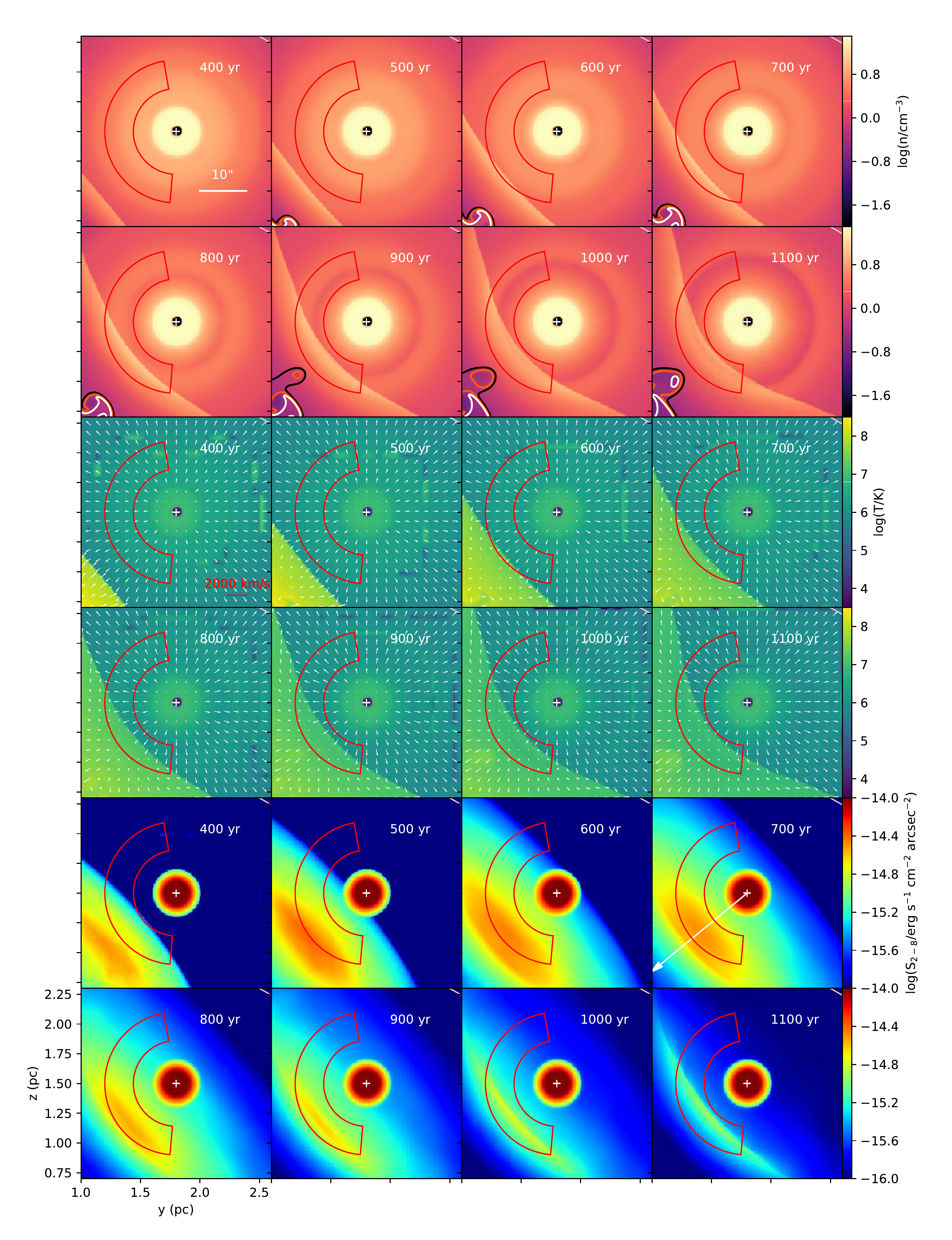}
\caption{Similar to Figure~\ref{fig:He_tot_long}, but with a higher metallicity (6 Z$_{\odot}$) for the ejecta of Sgr A East. }.
\label{fig:ZZ}
\end{figure*}

Comparing with {\it E5M1.3} (Figure~\ref{fig:He_tot_long}), it is found that in this simulation (Figure~\ref{fig:ZZ}) the overall evolution of the SNR and the formation of the ridge are rather insensitive to the metallicity.
The mean X-ray surface brightness of the ridge is $\rm 2.9\times10^{-15}~erg~s^{-1}~cm^{-2}~arcsec^{-2}$ at $t = 700$ yr, essentially identical to the case of \textit{E5M1.3}. This is understandable, because the ejecta has little contribution to the X-ray emission from the ridge even with the currently adopted high metallicity.

\section{The effect of a clumpy environment around the supernova explosion center}
\label{app:CL}
\begin{figure*}
\includegraphics[width=\textwidth]{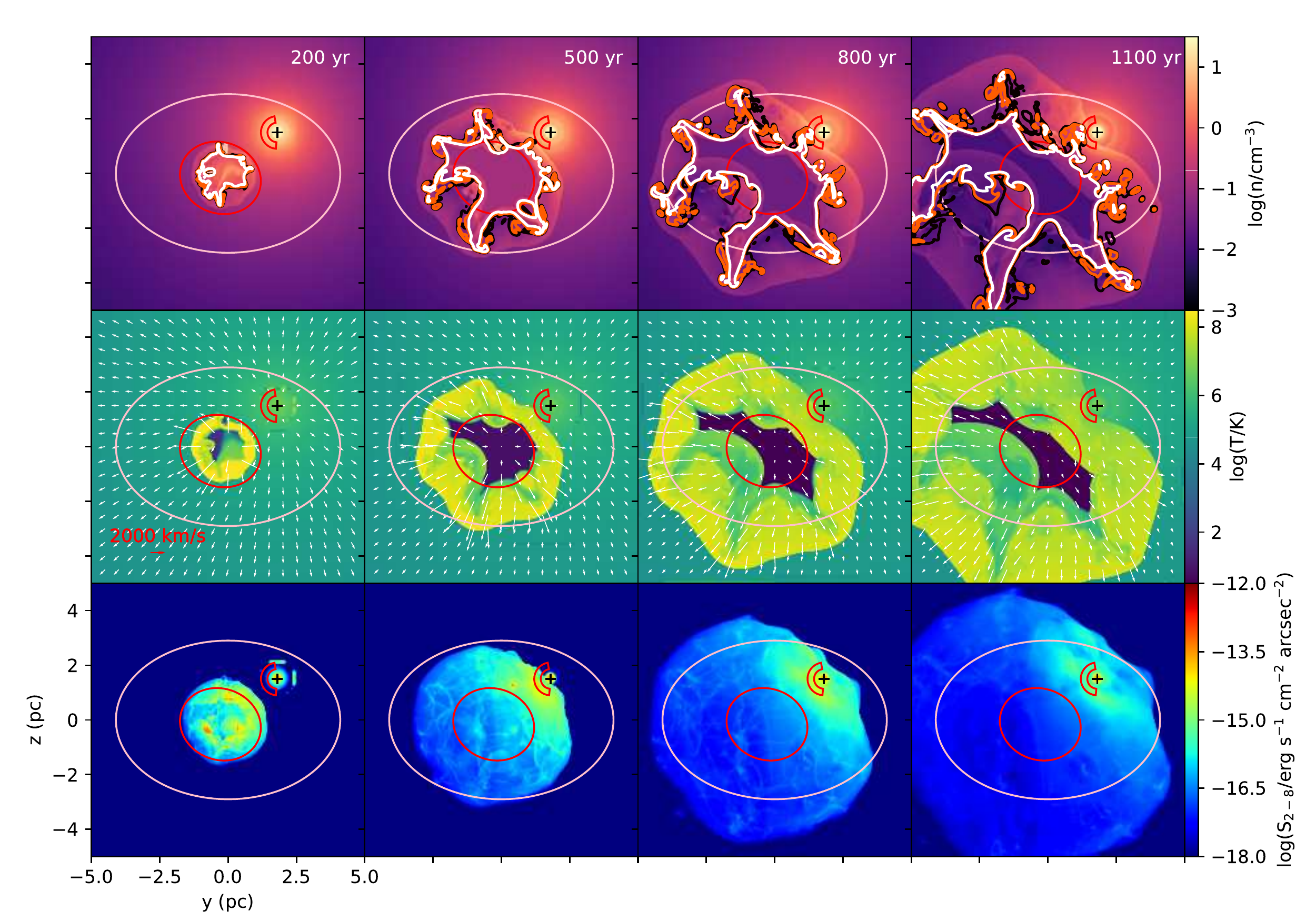}
\caption{Similar to Figure~\ref{fig:5}, but including some dense cloudlets surrounding the explosion center.}.
\label{fig:CL}
\end{figure*}

If there were only few small and low-density cloudlets around the explosion center, it would not significantly influence the evolution of Sgr A East and the formation of the X-ray ridge. However, if many large and dense cloudlets produced an envelope surrounding the explosion center, the results would be totally different. To illustrate this, we randomly add 10 cloudlets within the spherical region with a radius of 1.5 pc centered on the explosion center, and each cloudlet has a radius of 0.2 pc and a density of 10 cm$^{-3}$. The result is shown in Figure~\ref{fig:CL}. It can be seen that the resultant morphology is dramatically different from our standard simulations and also disagrees with the observation. Moreover, it cannot reproduce the “X-ray” ridge. In this regard, a clumpy environment around the explosion center is strongly disfavored. We note that the recent simulations of \citet{2022A&A...668A.124E} also included the influence of clouds, but they only considered the large {\it M50} cloud and limited physical conditions.

\end{document}